\documentclass[reprint,aps,prc,amsmath,amssymb,nofootinbib,superscriptaddress]{revtex4-2}

\usepackage{CJK}
\usepackage{graphicx}
\usepackage{dcolumn}
\usepackage{bm}
\usepackage{mathrsfs}   
\usepackage{color,xcolor}  
\usepackage{multirow}  
\usepackage{booktabs} 

\usepackage{hyperref}

\newcommand{\bbm}{\begin{bmatrix}}
\newcommand{\ebm}{\end{bmatrix}}
\newcommand{\bBm}{\begin{Bmatrix}}
\newcommand{\eBm}{\end{Bmatrix}}
\newcommand{\bpm}{\begin{pmatrix}}
\newcommand{\epm}{\end{pmatrix}}

\begin{document}


\title{Low-momentum relativistic nucleon-nucleon potentials I: Nuclear matter}

\author{Chencan Wang}
\email{wangchc5@mail.sysu.edu.cn}
\affiliation{School of Physics and Astronomy, Sun Sat-Sen University, Zhuhai 519082, China}


\author{Sibo Wang}
\affiliation{Department of Physics, Chongqing University, Chongqing 401331, China}

\author{Hui Tong}
\affiliation{Helmholtz-Institut f$\ddot{u}$r Strahlen- und Kernphysik and Bethe Center for Theoretical Physics, Universit$\ddot{a}$t Bonn, D-53115 Bonn, Germany}

\author{Jinniu Hu} 
\affiliation{School of Physics,  Nankai University, Tianjin, 300071, China}

\author{Jiangming Yao}  
\affiliation{School of Physics and Astronomy, Sun Sat-Sen University, Zhuhai 519082, China}

\begin{abstract} 

A series of relativistic one-boson-exchange potentials for two-nucleon system, denoted as OBEP$\Lambda$, is constructed with a momentum cutoff $\Lambda$ ranging from $\infty$ to 2 fm$^{-1}$. These potentials are developed by simultaneous fitting to nucleon-nucleon ($NN$) scattering phase shifts, low-energy scattering length, effective range, and the binding energy of the deuteron. The momentum-space matrix elements of the low-momentum OBEP$\Lambda$ ($\Lambda\leqslant 3$ fm$^{-1}$) demonstrate consistency with the universal behaviors observed in other realistic $NN$ potentials evolved by renormalization group methods. These OBEP$\Lambda$s are applied to calculate the equation of state of symmetric nuclear matter (SNM) within either the nonrelativistic (NR) Brueckner-Hartree-Fock (BHF) or relativistic Brueckner-Hartree-Fock (RBHF) frameworks. The results show that the saturation properties of SNM are reproduced  qualitatively from the RBHF calculation, but not from the NR-BHF calculation.
    This study highlights the relativistic mechanism in explaining
    the saturation properties of nuclear matter.
    The remaining discrepancy in reproducing empirical saturation properties in the RBHF calculation using the OBEP$\Lambda$s
    signals the necessity of including three-nucleon correlations or genuine three-nucleon forces. 
\end{abstract}
\date{\today} 
\maketitle

\section{INTRODUCTION}
\label{sec-intro}
    The nucleon-nucleon ($NN$) potential serves as a crucial input for nuclear \emph{ab initio} calculations. Originating from the 1960s, the meson-exchange model stands as an effective framework for deriving realistic $NN$ potentials~\cite{erkelenz1974pr,machleidt1987pr,machleidt1989anp}. Within this model, the one-boson exchange potentials (OBEPs), namely the Bonn potential and the highly accurate charge-dependent Bonn (CD-Bonn) potential, were proposed and remain frequently utilized in present-day \emph{ab initio} calculations~\cite{brockmann1990prc,machleidt2001prc}. The preservation of Dirac spinors and covariant operators in Bonn potentials enables the applicability to relativistic \emph{ab initio} approaches, such as the relativistic Brueckner-Hartree-Fock (RBHF) theory. However, the original CD-Bonn potential, utilizing the pseudoscalar (ps) type of pion-nucleon ($\pi N$) coupling, cannot be employed in RBHF calculations for nuclear matter due to the unphysically large self-energies induced by the ps $\pi N$ vertex~\cite{gross-beolting1999npa}. Yet, by substituting the ps $\pi N$ coupling with a pseudovector (pv) type, the modified CD-Bonn potentials become viable for relativistic applications~\cite{wang2020cpc,wang2020jpg}.

   Over the past two decades, substantial progress has been made in developing $NN$ potentials rooted in chiral effective field theory (chEFT) and renormalization group (RG) methods \cite{epelbaum1998plb,epelbaum1999npa,epelbaum2009rmp,bogner2010ppnp,machleidt2011pr}. Consequently, various nonrelativistic (NR) $NN$ potentials have been formulated at different resolution scales, characterized by specific momentum cutoffs $\Lambda$ \cite{machleidt2011pr,epelbaum2005npa,hebeler2011prc}. Additionally, three-nucleon force (3NF) with specified cutoff $\Lambda_{3N}$ arises naturally, either at the next-next-to-leading (N${}^2$LO) order in chEFT or through RG evolution in the flow equation \cite{epelabum2006ppnp,bogner2008annphys,hebeler2021pr}. In NR \emph{ab initio} calculations, the cutoff dependence of few-body observables directly reflects the residual many-body forces \cite{noggaprc2004,bogner2008npa}. Furthermore, variations in many-body observables concerning $\Lambda/\Lambda_{3N}$ provide insights into estimating theoretical uncertainties \cite{hebeler2011prc,tew2013prl,drischler2016prc,drischler2019prl}. This encourages the development of OBEPs with different momentum cutoffs $\Lambda$. Employing these potentials in relativistic \emph{ab initio} calculations of nuclear many-body systems, such as nuclear matter, enables the exploration of cutoff-dependent equation of state (EOS) from a relativistic perspective.


  Nuclear matter stands as a research topic of great interest in  nuclear physics since it helps us understand the bulk properties  of finite nuclei and the evolution of astrophysical objects like neutron stars.  In particular, the properties of nuclear matter around the saturation  density $n_0=0.16$ fm$^{-3}$ provide benchmarks to test the validity  of underlying $NN$ potentials and many-body methods~\cite{simonis2017prc,ekstroem2018prc,hoppe2019prc, sammarruca2020prc}.  In addition, the knowledge of the EOS of nuclear matter
    at supra-saturation densities is important to understand the 
    formation and structure of neutron stars
    ~\cite{lattimer2012arnps,oertel2017rmp,carreau2019prc,wang2020apj}, 
    as well as the particle production in the heavy-ion collision
    (HIC)~\cite{danielewicz2002sci,fevre2016npa,hermann2022ppnp}.  
    Early attempts to attack the problem were based on 
    nonperturbative approaches such as NR variational method or 
    Brueckner theory with traditional $NN$ potentials 
    \cite{day1978rmp,pandharipande1979rmp,clark1979ppnp, 
          day1979npa,jensen1995pr}. 
          
    It has been observed that the saturation properties of symmetric nuclear matter (SNM), derived from various $NN$ potentials, align within the ``Coester band'', systematically differing from the empirical saturation region \cite{coester1970prc,day1976prc,day1981prl,li2006prc}. This led to the conclusion that relying solely on $NN$ potentials fails to quantitatively replicate correct saturation properties. This highlighted the pivotal role of 3NF in understanding the saturation mechanism of nuclear matter \cite{day1981prl,bogner2005npa}. In contrast, the RBHF framework using only Bonn potentials, without explicit inclusion of 3NF, nearly reproduced SNM's saturation properties \cite{brockmann1990prc,gross-beolting1999npa}. RBHF theory hinges on two primary features: the effective Dirac spinor of nucleons, where the lower components rely on the effective Dirac mass, introducing additional density dependence in relativistic kinetics and the $G$-matrix; and the Lorentz structure of the self-energy, notably the attractive scalar self-energy and the repulsive timelike vector self-energy \cite{horowitz1987npa,sehn1997prc,fuchs1998prc}. The emergence of saturation properties in SNM results from the intricate balance between the linearly increasing vector self-energy and the gradually diminishing scalar self-energy \cite{ring1996pr}. Recent progressions in this field involve the successful application of RBHF theory, notably in fully self-consistent calculations for finite nuclei \cite{shen2016cpl,shen2017prc}, and nuclear matter calculations encompassing the complete Dirac space \cite{wang2021prc,wang2022prc}. 
    
    It's noteworthy that the availability of realistic $NN$ potentials suitable for RBHF calculations is severely limited, primarily confined to the three Bonn potentials developed more than thirty years ago~\cite{brockmann1990prc}. Additionally, these potentials lack specification regarding resolution scales, and the uncertainty assessment within the relativistic many-body method remains unexplored. To revitalize research in relativistic nuclear \emph{ab initio} calculations, our initial step involves constructing a series of OBEPs, explicitly incorporating momentum cutoffs $\Lambda$ (referred to as OBEP$\Lambda$s). These OBEP$\Lambda$s will be employed within the RBHF framework to compute the EOS of nuclear matter. 

    This paper is organized as follows. In Sec.~\ref{OBEPL-theory},
    the theoretical framework for OBEP, 
    the scattering equation, and the RBHF theory for nuclear matter 
    will be briefly reviewed.  
    In Sec.~\ref{OBEPL-resultA},
    the fitting protocol and the parameters of OBEP$\Lambda$s
    will be given, and the potential matrix elements 
    and the calculated $NN$ observables with OBEP$\Lambda$s will 
    also be provided. 
    In Sec.~\ref{OBEPL-resultB}, we will present
    nuclear matter results from both NR-BHF and RBHF 
    calculations with these OBEP$\Lambda$s, shedding light on the implications of relativity in the saturation mechanism of SNM. 
    Finally the summary and perspectives will be given in 
    Sec.~\ref{OBEPL-summary}.

\section{THEORETICAL FRAMEWORKS}~\label{OBEPL-theory}
\subsection{One-boson-exchange potential and NN observables}
Analogous to the Bonn potentials~\cite{brockmann1990prc}, present OBEP$\Lambda$s 
are developed based on the exchange of $\pi$, $\eta$ in pv coupling,  
$\sigma$, $\delta$ mesons in 
    scalar (s) coupling,  and $\omega$, $\rho$ mesons in vector 
    (v) coupling.  These nucleon-meson interaction Lagrangians are 
	\begin{subequations}\label{OBEPL-Lag}
	\begin{align} 
		\mathcal{L}^\mathrm{(pv)} &= 
            -\frac{f_\mathrm{pv}}{m_\mathrm{pv}}
		\bar{\psi}\gamma^5\gamma^\mu\psi \cdot 
		\partial_\mu  \phi^\mathrm{(pv)}, 
		\\
		\mathcal{L}^\mathrm{(s)}  & = 
		+g_\mathrm{s}\bar{\psi}\psi\cdot\phi^\mathrm{(s)}, 
		\\
		\mathcal{L}^\mathrm{(v)}  &=-g_\mathrm{v}
		\bar{\psi}\gamma^\mu\psi \cdot \phi_\mu^\mathrm{(v)} 
		-\frac{f_\mathrm{v}}{2M}\bar{\psi}\sigma^{\mu\nu}\psi
		\cdot \partial_\mu\phi_\nu^\mathrm{(v)}.
	\end{align}
	\end{subequations}
 The $NN$ potential in the center-of-mass (CM) frame is obtained from tree-level Feynman amplitude: 
	\begin{align}\nonumber 
		V(\mathbf{q}',\mathbf{q}) =&-\sum^{\mathrm{all~mes.}}_a
		\bar{u}_1(\mathbf{q}')\Gamma^{(1)}_a u_1(\mathbf{q})
		\frac{\mathcal{F}_a(Q^2)}{Q^2 + m_a^2}
            \\ \label{OBEPL-V}
            &  \times  \bar{u}_2(-\mathbf{q}')\Gamma^{(2)}_a u_2(-\mathbf{q}).
	\end{align} 
 Here, the subscript $a$ represents all the six mesons,  $\mathbf{q}$ and $\mathbf{q'}$ representing the incoming  and outgoing relative momenta, $u_i~(i=1,2)$ for nucleon spinor and $\Gamma^{(i)}_a$  for different meson-nucleon coupling vertices.  The $\mathbf{Q} = \mathbf{q}'-\mathbf{q}$ is three-momentum  transfer. The form factor   $\mathcal{F}_a(Q^2) =\exp[-(Q^2 + m_a^2)^2/\Lambda_a^4]$ 
        is used on the meson propagator to alter the behavior
        of local momentum transfer, 
        in which a meson-dependent parameter $\Lambda_a$ is 
        introduced.  This choice of form factor is tested to be
        more suitable in our fitting procedure, and different
        from $(\Lambda_a^2-m_a^2)^2/(\Lambda^2 +Q^2)^2$ used in the Bonn potentials.	
        
 In addition, we introduce the following non-local regulator
	\begin{equation}\label{OBEPL-RVR}
    	V_\Lambda(\mathbf{q}',\mathbf{q}) = 
    	\mathcal{R}(q') V(\mathbf{q}',\mathbf{q})
    	\mathcal{R}(q),
	\end{equation} 
        with 
        \begin{equation}\label{OBEPL-Reg}
            \mathcal{R}(q) = \exp[-(q^{2n}/\Lambda^{2n})].
        \end{equation} 
 The regulators above strongly suppress the matrix elements with relative momenta larger than the cutoff $\Lambda$. 

The $T$ matrix for the $NN$ scattering process is obtained by the Thompson  equation~\cite{brockmann1990prc}:
	\begin{align}\nonumber 
		T(\mathbf{q}',\mathbf{q};W_q)=&
		V_\Lambda(\mathbf{q}',\mathbf{q})
		+\int \frac{\mathrm{d}^3 p}{(2\pi)^3}
		\frac{M^2} {E^2_p}V_\Lambda
		(\mathbf{q}',\mathbf{p})
            \\ \label{OBEPL-ThompEq} 
            \times&  \frac{1}{W_q - W_p+i\varepsilon} 
		T(\mathbf{p},\mathbf{q};W_q),
	\end{align}
	where $E_p= \sqrt{\mathbf{p}^2 + M^2}$ is the nucleon  
	on-shell energy. $W_q = 2E_q$ and $W_p=2E_p$ 
	are the initial and intermediate two-nucleon 
	energy in CM frame respectively. 
	The partial-wave scattering matrix is obtained by 
	\begin{equation}\label{OBEPL-Smat}
		\mathcal{S}_{\ell' \ell} = \delta_{\ell'\ell}
		- i \pi  \frac{ q M^2}{E_q}\langle \ell' s j|
		T(W_q)|\ell s j\rangle.
	\end{equation}
	Corresponding phase shifts $\delta$ in uncoupled channel
	are given by  $\mathcal{S}_{\ell \ell} = e^{i2\delta_\ell}$.
   	For coupled channels, the phase shifts $\delta_{\ell\pm 1}$
   	and mixing angle $\varepsilon_j$ are obtained by 
   	\begin{equation}\label{OBEPL-Smat-coup}
  			\begin{pmatrix} 
  				\mathcal{S}_{--} & \mathcal{S}_{-+}\\
  			 	\mathcal{S}_{+-} & \mathcal{S}_{++} 
  			\end{pmatrix}
  		 	=\begin{pmatrix}
  		      \cos2\varepsilon_j e^{2i\delta_-} & 
  				i\sin2\varepsilon_j e^{i(\delta_-+\delta_+)}  \\
  				i\sin2\varepsilon_j e^{i(\delta_-+\delta_+)} 
  				& \cos2\varepsilon_j e^{2i\delta_+}
  			\end{pmatrix},
   	\end{equation}
   	where $\pm$ stand for $\ell = j\pm 1$.
    	
    The binding energy $E_d$ and wave functions 
    $(\psi_S,\psi_D)^T$ of deuteron are obtained  by solving the following homogeneous Thompson equation
    \begin{align}\nonumber \label{OBEPL-Deut}
        \begin{pmatrix} 
            \psi_S(q) \\  \psi_D(q) 
        \end{pmatrix} =& 
        \frac{1}{2M-E_d-W_q}
        \int_0^{+\infty} p^2\mathrm{d}p
            \frac{M^2} {E_p^2} \\
            \times & \begin{pmatrix} 
            V_{\Lambda,SS}(q,p)& V_{\Lambda,SD}(q,p)  \\
            V_{\Lambda,DS}(q,p)& V_{\Lambda,DD}(q,p) 
        \end{pmatrix}
        \begin{pmatrix}
            \psi_S(p) \\ \psi_D(p)
        \end{pmatrix}.
    \end{align}
\subsection{The RBHF theory with projection method}
    The single-nucleon motion in nuclear matter 
    follows the Dirac equation 
    \begin{equation}\label{OBEPL-DiracEq}
        [\bm{\alpha}\cdot \mathbf{k} + \beta M +\beta \Sigma(k)]
        u(\mathbf{k},\lambda) = E_k u(\mathbf{k},\lambda),
    \end{equation}
    The self-energy in nuclear matter can be expressed as $\Sigma = \Sigma_\mathrm{S} - \gamma^0 \Sigma_0 + \bm{\gamma}\cdot \mathbf{k} \Sigma_\mathrm{V}$, where $\Sigma_\mathrm{S}$, $\Sigma_0$, and $\Sigma_\mathrm{V}$ represent the scalar self-energy, time-like and space-like vector self-energies, respectively. Here, $\lambda=\pm 1/2$ denotes helicity.
    
    With the definitions of reduced Dirac mass and effective energy  
    \begin{equation}\label{OBEPL-effME}
        M^* = \frac{M + \Sigma_\mathrm{S}}{1+\Sigma_\mathrm{V}},
        \quad E^*_k = \frac{E_k - \Sigma_0}{1+\Sigma_\mathrm{V}},
    \end{equation}		   
    the solutions to the Dirac equation are     
    $E^*_k= \sqrt{\mathbf{k}^2 + M^{*2}}$
    and plane-wave spinor 
    \begin{equation}
        u(\mathbf{k},\lambda) = \sqrt{\frac{E^*_k + M^*}{2M^*}}
        \begin{pmatrix} 
            1  \\ \frac{\bm{\sigma}\cdot\mathbf{k}}{M^* + E^*_k}
        \end{pmatrix}
        |\lambda\rangle.
    \end{equation}
    
    The effective $NN$ potential in nuclear matter is obtained with the Brueckner $G$-matrix:
    \begin{align}\nonumber 
        G(\mathbf{q}',\mathbf{q}|\mathbf{P},W_q)&
        =V_\Lambda(\mathbf{q}',\mathbf{q})
        +\int\frac{\mathrm{d}^3 p}{(2\pi)^3}
        \frac{M^{*2}}{E^{*2}_p}V_\Lambda
        (\mathbf{q}',\mathbf{p}) 
        \\ \label{OBEPL-ThompQEq}
        &\times \frac{Q(\mathbf{p},\mathbf{P})}
        {W_q- W_p+i\varepsilon} 
        G(\mathbf{p},\mathbf{q}|\mathbf{P},W_q),
    \end{align}
    where $\mathbf{P}$ is the CM momentum and   
    $Q$ is Pauli blocking operator prohibiting 
    nucleon scattering into occupied states.
 
    As illustrated in Ref.~\cite{gross-beolting1999npa}, the scalar and time-component vector self-energies may exhibit unphysically large values due to the inadequate treatment of the one-pion-exchange potential $V_\pi$. To mitigate this issue, the subtracted $T$-matrix scheme is proposed, wherein the $G$-matrix is split into two components, $G=V_\mathrm{pv}+\Delta G$, and $V_\mathrm{pv}=V_\pi + V_\eta$. The transformation of $\Delta G$ from the $\ell s j$ representation to partial-wave helicity representation enables the derivation of invariant amplitudes $F$.
    \begin{equation}\label{OBEPL-DG-prj}
    \Delta G = F_\mathrm{S}\Gamma_\mathrm{S} + 
        F_\mathrm{V}\Gamma_\mathrm{V}+ 
    F_\mathrm{T}\Gamma_\mathrm{T} + 
        F_\mathrm{P}\Gamma_\mathrm{P}+
    F_\mathrm{A}\Gamma_\mathrm{A},   
    \end{equation}
    with the pseudoscalar-type covariant basis
    \begin{subequations}
    \begin{align}\label{OBEPL-PSs}  
        \Gamma_\mathrm{S} &= 1_1\otimes 1_2, \\
        \label{OBEPL-PSv}
        \Gamma_\mathrm{V} &=(\gamma^\mu)_1\otimes 
        (\gamma_{\mu})_2, \\ 
        \label{OBEPL-PSt}
        \Gamma_\mathrm{T} &=
        (\sigma^{\mu\nu})_1\otimes(\sigma_{\mu\nu})_2, 
        \\   
        \label{OBEPL-PSa}
        \Gamma_\mathrm{A} &= (\gamma^5\gamma^\mu)_1
        \otimes (\gamma^5\gamma_\mu)_2, \\ 
        \label{OBEPL-PSp}
        \Gamma_\mathrm{P} &=(\gamma^5)_1\otimes (\gamma^5)_2.
    \end{align}
    \end{subequations} 
    The self-energies generated by $\Delta G$ can be calculated by 
    \begin{subequations}\label{OBEPL-SE}
    \begin{align}\label{OBEPL-SESs}
    \Sigma_\mathrm{S}(k) &= \int \frac{\mathrm{d}^3 k'}{(2\pi)^3}
    \frac{M^*}{E_{k'}^*} F_\mathrm{S}(q,q), 
    \\
    \label{OBEPL-SESo}
    \Sigma_0(k)& = -\int \frac{\mathrm{d}^3 k'}{(2\pi)^3}
    F_\mathrm{V}(q,q), 
    \\
    \label{OBEPL-SESv}
    \Sigma_\mathrm{V}(k) &=- \frac{1}{k^2}
    \int \frac{\mathrm{d}^3 k'}{(2\pi)^3}
    \frac{\mathbf{k}\cdot \mathbf{k}'}{E_{k'}^*} 
    F_\mathrm{V}(q,q), 
    \end{align}
    \end{subequations}
    where $q=\frac{1}{2}\sqrt{s^* - 4M^{*2}}$	
    is the relative momentum in two-nucleon CM frame,
    $s^*=(E_k^*+E_{k'}^*)^2 - (\mathbf{k}+\mathbf{k}')^2$.
    
    The remaining $V_\mathrm{pv}$ is decomposed in 
    complete pseudovector representation, corresponding 
    formulae for self-energies are complicated, see 
    Ref.~\cite{gross-beolting1999npa} for details. Numerical computations demonstrate that the contributions from $V_\mathrm{pv}$ to self-energies are approximately one order of magnitude smaller than those originating from $\Delta G$.
    
    The Brueckner G-matrix~\eqref{OBEPL-ThompQEq} is solved 
    by iteration. After convergence, the binding energy per-nucleon in nuclear matter is given by $E/A = E_\mathrm{kin}/A + E_\mathrm{pot}/A$,
    where the kinetic term is calculated by
     \begin{equation}\label{OBEPL-EkinR}
        E_\mathrm{kin}/A =  
        \frac{1}{n} \sum_\lambda \int^{k_F} 
        \frac{\mathrm{d}^3k}{(2\pi)^3}  
        \langle	\bar{u}(\mathbf{k},\lambda)| \bm{\gamma}\cdot 
        \mathbf{k}+ M|u(\mathbf{k},\lambda)\rangle -M,
    \end{equation}
     with $n$ representing the nucleon number density. The average potential energy is given by
    \begin{align}  
        \nonumber 
        E_\mathrm{pot}/A 
        &=  \frac{1}{n} \sum_\lambda \int^{k_F} 
        \frac{\mathrm{d}^3k}{(2\pi)^3}  
        \langle	\bar{u}(\mathbf{k},\lambda)|
        \\        \label{OBEPL-EpotR}
        &
        \times (M^*\Sigma_\mathrm{S}/E^*_k-\Sigma_0+ k^2\Sigma_\mathrm{V}/E^*_k)
        |u(\mathbf{k},\lambda)\rangle. 
    \end{align} 

\section{RESULTS AND DISCUSSIONS}
\subsection{The fitting procedure and NN observables}\label{OBEPL-resultA}

    \begin{table*}
    \tabcolsep=8pt
    \centering
    \caption{Mesons parameters in OBEP$\Lambda$s with the 
    variation of cutoffs $\Lambda$ (in unit fm$^{-1}$).
    In each row, the coupling parameters are given  
    as $g_a~(\Lambda_a)$.  
    The meson masses (in the parenthesis after each meson) 
    and $\Lambda_a$ are given in unit GeV.
    For pion and eta meson, we take $g_a = 2M f_a/m_a$,
    where the mass of nucleon is the averaged 
    value  $M=$938.919 MeV. 
    $\rho$ meson coupling constants are given 
    in the form of $g_\rho$--$f_\rho/g_\rho$.}
    \label{tab:mes_param}
    \begin{tabular}{ccccccc}	
        \hline \hline 
        $\Lambda$  & $\pi$ (0.138)& $\omega$ (0.783) 
        & $\rho$ (0.770) & $\eta$ (0.548) & $\delta$ (0.983)
        & $\sigma$ (0.550) 
        \\ \hline 
        $\infty$  
        &12.71 (0.97) &14.81 (1.05) & 2.38--6.0 (0.90) & 
        1.29 (0.77)   & 9.02 (0.96) &10.28 (1.83)\\
        5  & 12.69 (0.97) & 14.28 (1.03) & 2.16--6.8 (0.97) 
        & 4.59 (0.77) & 8.07 (0.95)  & 10.27 (1.75) 
        \\     
        4 & 12.66 (0.93) &13.55 (1.00) & 2.23--6.9 (1.01) 
        & 5.20 (0.77) & 8.80 (0.87) &  9.95 (1.50) \\ 
        3 & 12.67 (0.91)&12.98 (0.97)& 2.42--6.8 (1.03) 
        &  5.06 (0.77) & 10.31 (0.81) &9.70 (1.32)\\ 
        2 & 13.20 (0.90) & 12.15 (0.94)& 2.89--5.7 (1.04)
        & 0.00 (0.77) & 10.38 (0.78) &9.32 (1.23) \\
        \hline  \hline 
    \end{tabular}
    \end{table*}

 The parameters in the OBEP$\Lambda$ in Eq.~\eqref{OBEPL-V}  are determined by minimizing the objective function
     \begin{align}\nonumber
        f(\mathbf{X})  & = \sum_{\delta}   w_\delta^2
        \left(\delta  - \delta_\mathrm{NPWA} \right)^2 
        +w_d^2\left(E_d-E_d^\mathrm{(expt)}\right)^2 
        \\  \label{OBEPL-X2}
        & + \sum_{S}\left[w_a^2\left(a_S -a_S^\mathrm{(expt)}
        \right)^2 + w_r^2 \left(r_S-r_S^\mathrm{(expt)}
        \right)^2\right],
     \end{align}
    where $\mathbf{X}$ contains 12 variables, including
    $g_a~(f_a)$ and $\Lambda_a$ for the six mesons. 
    The neutron-proton (np) phase shifts of partial wave 
    $j\leqslant 4$  with laboratory energy  
    $E_\mathrm{lab}\leqslant 300$ MeV are calculated to 
    compare with phase shifts from Nijmegen 
    partial-wave analyse (NPWA)~\cite{stoks1994prc}.
    $E_d^\mathrm{(expt)} = 2.2246$~MeV is the binding energy
    of deuteron.  The low-energy scattering observables
    \cite{dumbrajs1983npb}, namely the
    $S$--wave scattering lengths
    $a_{^1\! S_0}^\mathrm{(expt)}=-23.75$ fm, 
    $a_{^3\! S_1}^\mathrm{(expt)}=5.42$ fm,
    and effective ranges
    $r_{^1 \! S_0}^\mathrm{(expt)} = 2.75$  fm,
    $r_{^3 \! S_1}^\mathrm{(expt)} = 1.76$  fm
    are also included in the fitting.
    $w_\delta$, $w_d$, $w_a$, $w_r$ are the weighting factors, 
    we employ $w_\delta=1/(\Delta \delta_\mathrm{NPWA})$,
    with $\Delta \delta_\mathrm{NPWA}$ being the phase shift 
    uncertainty in NPWA. 
    We employ $w_d=1000$ and $w_r,~w_a=100$ to ensure both 
    partial-wave phase shifts and low-energy observables 
    can be simultaneously reproduced.

\begin{figure} 
        \centering
        \includegraphics[width=\columnwidth]
        {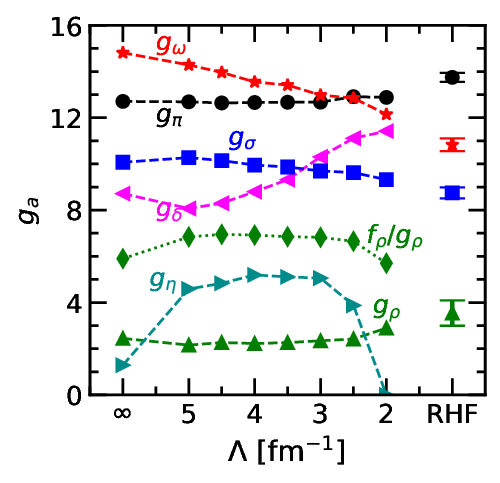}
        \caption{The meson-nucleon coupling constants $g_a$ running with
        external relative momentum cutoff $\Lambda$. The error bars are
        statistic mean values and standard deviations of
        meson-nucleon coupling parameters for the RHF approaches~\cite{long2006plb,long2007prc}.}
        \label{fig:OBEPL-grunL}
    \end{figure}

   During the fitting, $n=3$ is applied in the regulator~\eqref{OBEPL-RVR} for $\Lambda\geqslant$ 4 fm$^{-1}$, and $n=4$ is used for $\Lambda \leqslant$ 3 fm$^{-1}$. Slightly different parameterizations for the six mesons are allowed to minimize $f(\mathbf{X})$ at each cutoff, but the nuclear matter results are insensitive to such small changes in parameters. The mesons' parameters are finally determined at each $\Lambda$ by both minimizing $f(\mathbf{X})$ and considering the continuity between adjacent cutoffs. These parameters are listed in Tab.~\ref{tab:mes_param}. In Fig.~\ref{fig:OBEPL-grunL}, variations of meson-nucleon coupling constants with respect to decreasing cutoffs are plotted.
  As shown in the figure, $g_\pi$, $g_\rho$, and $f_\rho/g_\rho$ only vary a little, while $g_\sigma$ and $g_\omega$ are monotonously decreasing with $\Lambda$.  The $g_\eta$ and $g_\delta$ show nonmonotonic behaviors, and the $\eta$ meson finally turns out to be redundant at $\Lambda =2$ fm$^{-1}$ according to the minimization procedure. For comparison, meson-nucleon coupling constants from relativistic Hartree-Fock (RHF) density functional parameter sets, such as PKA1, PKO$i$ ($i=1,2,3$) \cite{long2006plb,long2007prc}, are also plotted in the form of error bars, according to their statistical mean values and standard deviations. 
  It can be found that present coupling constants from OBEP$\Lambda$ have the tendencies to close those of RHF model at low cutoff.

    \begin{figure*} 
        \centering
        \includegraphics[width=1.8\columnwidth]
        {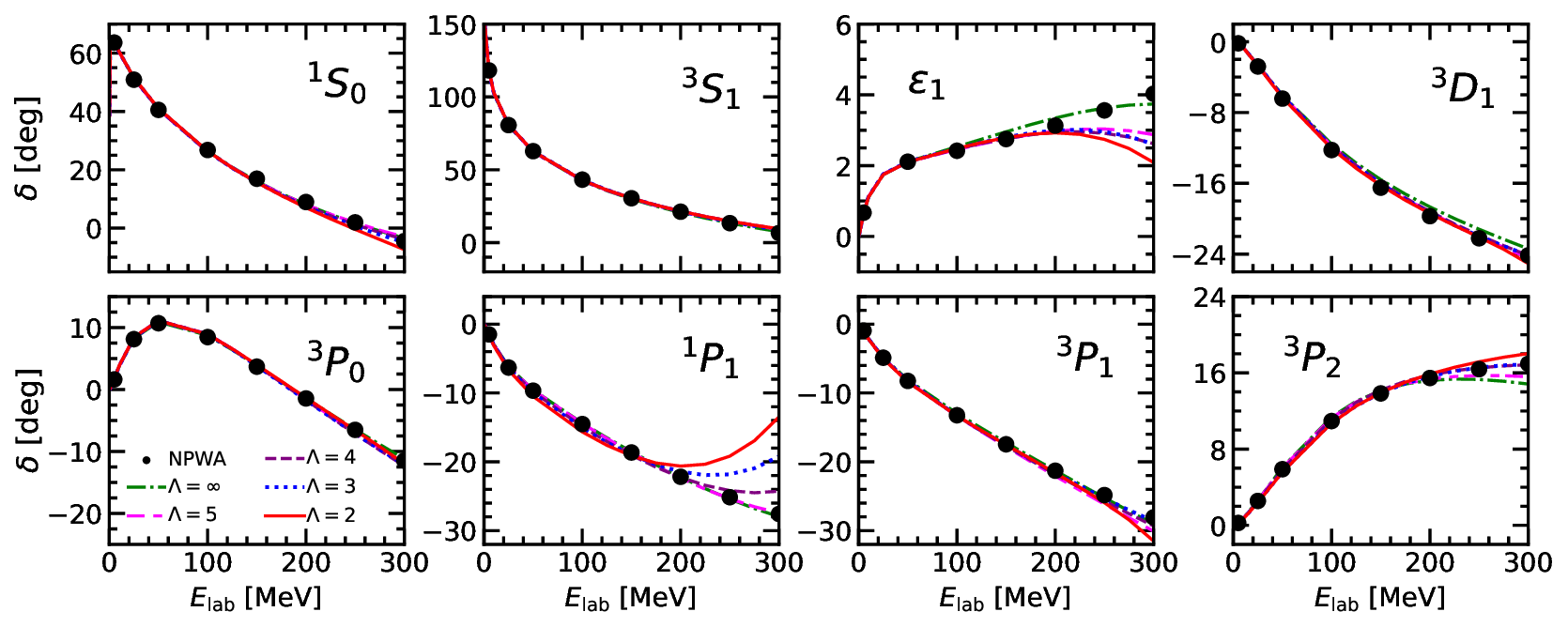}
        \caption{Neutron-proton partial-wave phase shifts $j\leqslant 2$ calculated with the OBEP$\Lambda$s.}
        \label{fig:OBEPL-phshft}
    \end{figure*}

    The  phase shifts with $j\leqslant2$, calculated using OBEP$\Lambda$s and depicted in Fig.~\ref{fig:OBEPL-phshft}, demonstrate good 
    agreement with the NPWA analysis~\cite{stoks1994prc} up to $E_\mathrm{lab} = 200$ MeV. This congruence extends to the peripheral partial waves in our calculations. However, for $E_\mathrm{lab} > 200$ MeV, discernible deviations arise, particularly noticeable in the ${}^1 P_1$ channel. Although this channel is specially treated in both Bonn potentials and the CD-Bonn potential,  in our work, there are no specific partial-wave refinements in . Tab.~\ref{tab:OBEPL-deut} presents the deuteron properties computed using OBEP$\Lambda$s. Remarkably, properties such as the matter radius $r_d$, quadrupole momentum $Q_d$, and asymptotic $D/S$ ratio remain largely invariant despite variations in cutoffs, while the $D$-state probability $P_D$ gradually diminishes as $\Lambda$ decreases. 
    $P_D$ is related to the tensor components of $NN$ interaction, when the cutoffs decreases, the tensor components are suppressed.  
    This reduction in $P_D$ reflects an understanding that it is not a directly observable; its dependence on the cutoff is also observed in the RG evolution as documented in \cite{bogner2007npa}.
    
    \begin{table}
    \centering
    \tabcolsep=8pt
        \caption{Deuteron properties predicted by OBEP$\Lambda$s.
        $r_d$ is the matter radius,
        $Q_d$ is the quadrupole momentum, 
        $D/S$ is the ratio of asymptotic $D$ and $S$ 
        state amplitudes, and $P_D$ is the 
        $D$--state probability.} 
        \begin{tabular}{ccccc}
        \hline \hline 
        $\Lambda$  & $r_d$ [fm] 
        & $Q_d$ [fm$^2$] & $D/S$ & $P_D$ (\%)  \\
        \hline
        $\infty$ & 1.967 & 0.2634 & 0.0247 &5.498 \\ 
              5  & 1.967 & 0.2630 & 0.0246 &5.429 \\ 
              4  & 1.964 & 0.2625 & 0.0246 &5.205 \\
              3  & 1.964 & 0.2634 & 0.0246 &4.827 \\
              2  & 1.967 & 0.2634 & 0.0248 &4.246\\
            expt. & 1.975 & 0.2859 & 0.0256& \\		
        \hline \hline 
        \end{tabular}
        \label{tab:OBEPL-deut}
    \end{table}

\begin{figure} 
        \centering
        \includegraphics[width=\columnwidth]
        {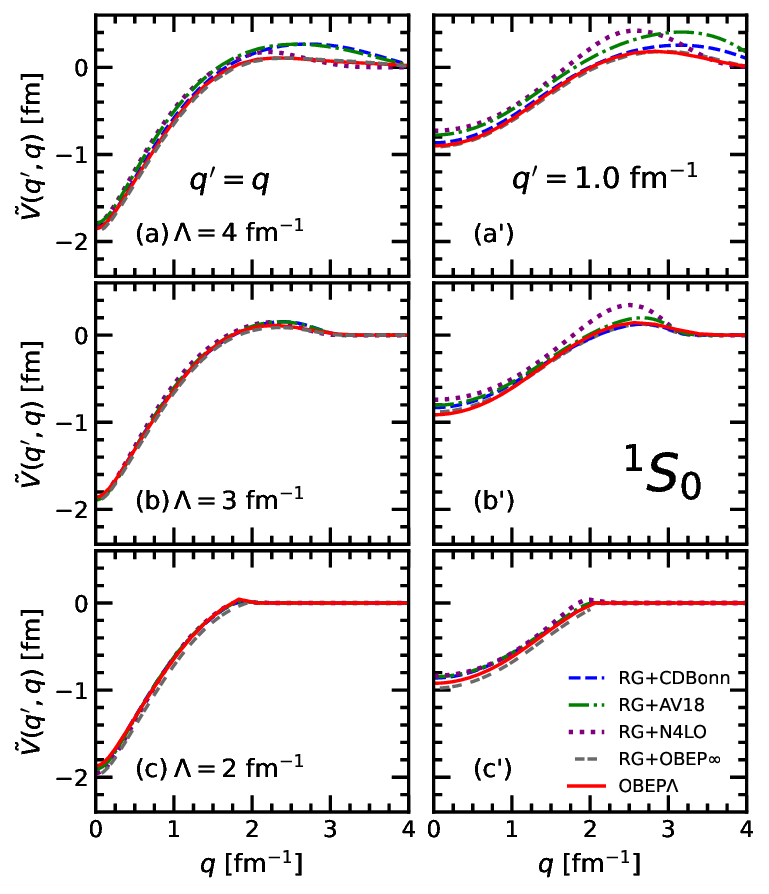}
        \caption{The matrix elements of on-shell (left panels) and half-on-shell  (right panels) OBEP$\Lambda$  potentials in  the ${}^1S_0$ channel, in comparision with other RG evolved 
        realistic $NN$ potentials, including CD-Bonn, AV18 and N${}^4$LO.
        Here we use the same convention as Ref.~\cite{bogner2010ppnp}, 
        $\tilde{V}(q',q) = \pi M V(q',q)/2$.}
        \label{fig:OBEPL-compare}
    \end{figure}

  Figure~\ref{fig:OBEPL-compare} displays a comparison of momentum-space matrix elements in the ${}^1S_0$ channels of OBEP$\Lambda$s ($\Lambda\leqslant4$ fm$^{-1}$) with other realistic $NN$ potentials, including OBEP$\infty$, the CD-Bonn potential \cite{machleidt2001prc}, Argonne $v18$ (AV18) \cite{wiringa1995prc}, and the chiral nuclear force at the fifth order (with the original cutoff $\Lambda_\chi=500$ MeV, denoted as N${}^4$LO) evolved by a smooth RG technique \cite{entem2017prc,bogner2007npa}. In the left panels (a), (b), and (c), on-shell ${}^1S_0$ potential elements are provided. These panels reveal that as $\Lambda$ decreases to $\Lambda\leqslant3$ fm$^{-1}$, the on-shell potential matrix elements of CD-Bonn, AV18, N${}^4$LO, and our OBEP$\Lambda$ converge closely. The corresponding half-on-shell potential matrix elements in the right panels (a'), (b'), and (c') generally mirror the trends observed in the on-shell cases. Similar behavior is noted in potential matrix elements of other partial-wave channels. Notably, the potential matrix elements of OBEP$\Lambda$s closely resemble those of other realistic potentials after RG evolution down to $\Lambda=$ 2 fm$^{-1}$, clearly demonstrating the universality of phase-shift equivalent $NN$ potentials \cite{bogner2010ppnp}.
  
    
\subsection{Nuclear matter results}
\label{OBEPL-resultB}

 \begin{figure}[bt]
        \centering
        \includegraphics[width=\columnwidth]{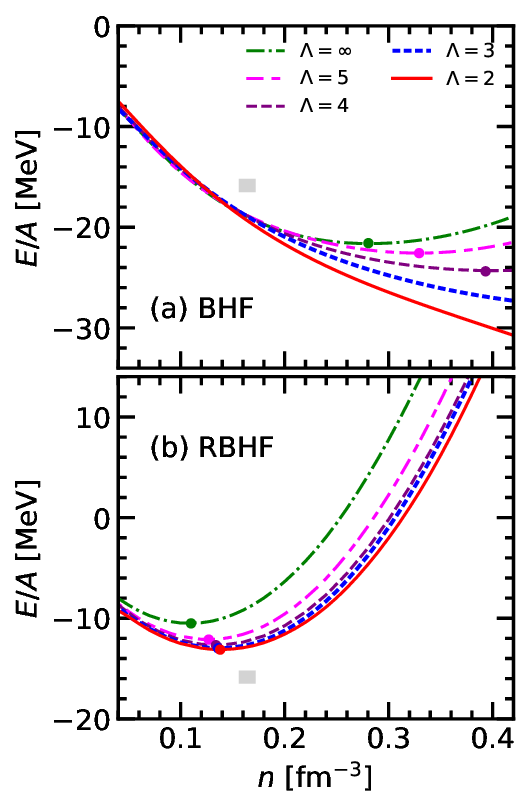}
        \caption{The energy per nucleon for the symmetric nuclear matter calculated by BHF and  RBHF approaches with different OBEP$\Lambda$s. The shaded area  indicates the empirical saturation region  with density $n=0.164\pm0.007$ fm$^{-3}$
            and $E/A=-15.86\pm 0.57$ MeV \cite{drischler2019prl}. The 
            energy minimum of each curve is marked by a dot.}
        \label{fig:snm_compare}
    \end{figure}

        \begin{figure}[bt]
        \centering
        \includegraphics[width=\columnwidth]{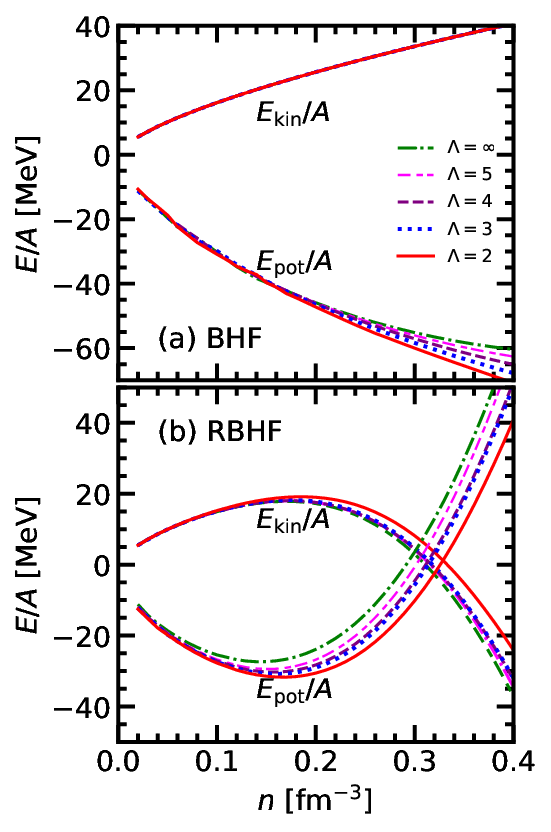}
        \caption{The kinetic and potential terms obtained by 
         BHF calculation in panel (a) and RBHF  calculation
         in panel (b), with OBEP$\Lambda$s as input.}
        \label{fig:ekin&epot}
    \end{figure}
    
     The OBEP$\Lambda$s potentials are utilized in nuclear matter calculations employing both NR-BHF and RBHF approaches. The resulting EOSs are depicted in Fig.~\ref{fig:snm_compare}, with the saturation points of each curve indicated. In the NR calculations, OBEP$\Lambda$s are augmented with minimal relativity \cite{brown1969npa}, 
    \begin{equation}
        V_{\Lambda,\mathrm{NR}}(\mathbf{q'},\mathbf{q}) = 
        \sqrt{\frac{M}{E_{q'}}}V_\Lambda(\mathbf{q'},\mathbf{q})
        \sqrt{\frac{M}{E_q}},
    \end{equation}
    while BHF calculations are executed under the continuous choice method \cite{baldo1991prc}. 

    In panel (a) of Fig.\ref{fig:snm_compare}, the NR-BHF results exhibit convergence before reaching the empirical saturation region, although they notably display an overbound nature. These findings qualitatively align with the outcomes from Refs.~\cite{bogner2007npa,hebeler2011prc}, particularly in instances where, at low cutoffs, the bare potentials demonstrate quantitative proximity, as depicted in Fig.\ref{fig:OBEPL-compare}. However, as densities surpass the empirical saturation region, the divergence of NR EOSs becomes evident. The extent of overbinding corresponds to the chosen cutoffs; notably, for cutoffs lower than 3 fm$^{-1}$, no saturation points are observed in the region $n\leqslant4$ fm${}^{-3}$. This absence underscores the necessity of incorporating 3NFs to elucidate the saturation mechanism for softer $NN$ potentials in NR frameworks.

    In panel (b), the situation regarding relativistic results differs notably. All RBHF calculations conducted with OBEP$\Lambda$s manifest saturation phenomena. However, the binding energy and density at each saturation point do not align adequately with those observed in the empirical saturation region, even with the softest potential at $\Lambda=2$ fm$^{-1}$. The convergence patterns of resulting EOSs in relation to cutoff variation also diverge from their non-relativistic counterparts; notably, the gaps between adjacent cutoffs decrease as the cutoff decreases. The disparity between current RBHF calculations and empirical saturation properties might possibly be attributed to unaccounted relativistic 3NF, or the exclusion of higher-order contributions in Bethe-Brueckner-Goldstone (BBG) expansion,
    which warrants exploration in future studies. Additionally, it's important to note that while the original Bonn A potential can nearly reproduce saturation properties \cite{brockmann1990prc}, but some of its phase shift predictions, especially the mixing parameter $\epsilon_1$ in the ${}^3S_1$--${}^3D_1$ channel, deviate considerably from NPWA even at very small
    $E_\mathrm{lab}$. In contrast, all the OBEP$\Lambda$s successfully reproduce $\varepsilon_1$ up to $E_\mathrm{lab}=200$ MeV.

       \begin{figure}[bt]
        \centering
        \includegraphics[width=\columnwidth]{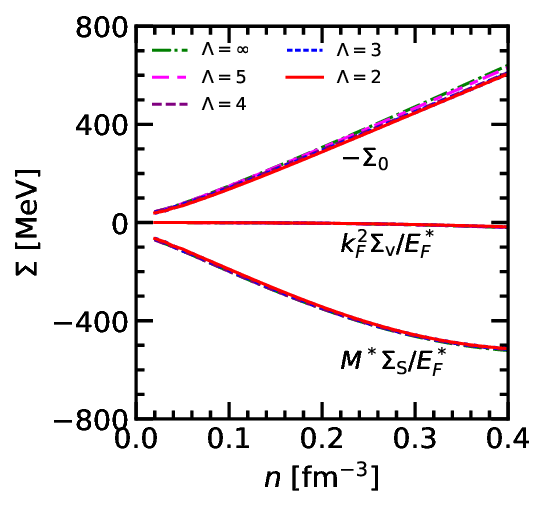}
        \caption{The self-energy components at Fermi 
        momentum $k_F$ obtained in RBHF calculations with 
        OBEP$\Lambda$s.}
        \label{fig:sef}
    \end{figure}

     \begin{figure}[]
        \centering
        \includegraphics[width=\columnwidth]{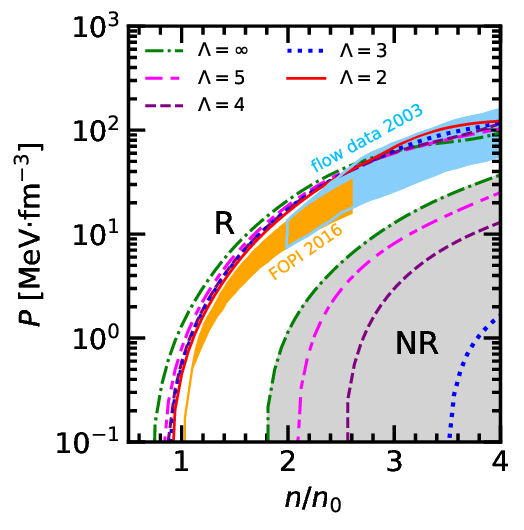}
        \caption{The pressure of symmetric nuclear matter 
        at zero temperature obtained by BHF and RBHF calcualtions
        with  OBEP$\Lambda$s (to distinguish, the NR-BHF results 
        are shaded in grey). 
        The shaded areas in orange and cyan 
        are experimental constraints from HIC experiments.}
        \label{fig:pres}
    \end{figure}

    To clarify the differences between BHF and RBHF results with
    the OBEP$\Lambda$s, we show in Fig.~\ref{fig:ekin&epot} the kinetic 
    and potential terms obtained from BHF and RBHF calculations
    respectively. For BHF calculation in panel (a), 
    the kinetic terms are  all the same as free Fermi gas,
    while the potential terms vary with cutoffs, which are 
    in correspondance with the divergence of EOSs shown in panel~(a)
    of Fig.~\ref{fig:snm_compare}. For RBHF calculations in
    panel (b), as Eq.~\eqref{OBEPL-EkinR} indicates, the kinetic 
    mass equals to the expectation value of relativistic kinetic
    operator $\bm{\gamma}\cdot \mathbf{k} + M$ minus rest mass.
    Before empirical saturation density, the two kinetic 
    contributions are close.
    Since Dirac mass appears in the lower component of 
    spinor $u(\mathbf{k},\lambda)$, the expectation 
    value of relativistic kinetic operator can be even smaller 
    than the rest nucleon mass at large densities. 
    
    The relativistic potential terms from Eq.~\eqref{OBEPL-EpotR}, 
    plotted in panel (b) of 
    Fig.~\ref{fig:ekin&epot} gain considerable repulsion 
    as compared with BHF results. 
    To understand the source of repulsion in relativistic 
    potential contributions, we present in Fig.~\ref{fig:sef} 
    the self-energy components appearing in Eq.~\eqref{OBEPL-EpotR}.
    Since $\Sigma_\mathrm{V}$ are one order of magnitude smaller than 
    $\Sigma_\mathrm{0}$ and $\Sigma_\mathrm{S}$,
    we will mainly focus on $\Sigma_\mathrm{0}$ and $\Sigma_\mathrm{S}$.
    By relativistic decomposition of $\Delta G$ as 
    Eq.~\eqref{OBEPL-DG-prj} and $V_\mathrm{pv}$ in 
    complete pv representation, both attractive $\Sigma_\mathrm{S}$
    and repulsive $-\Sigma_0$ as large to several hundreds 
    are generated for all cutoffs. The attractive 
    $\Sigma_\mathrm{S}$ gets quenched at large densities, 
    due to a factor $M^*/E_{k'}^*$ present in the integrand 
    of Eq.~\eqref{OBEPL-SESs}, while $-\Sigma_0$  increases almost 
    linearly  with increasing density. The cancellation
    between $M^*\Sigma_\mathrm{S}/E_k^*$ and  $-\Sigma_0$
    finally leads to  considerable repulsion in 
    Eq.~\eqref{OBEPL-EpotR} compared to BHF calculations.

    At zero temperature, the pressures of symmetric nuclear matter, given by $P=n^2 \frac{\partial(E/A)}{\partial n}$, are computed via both BHF and RBHF calculations utilizing OBEP$\Lambda$s as functions of density. Illustrated in Fig.~\ref{fig:pres}, the shaded areas represent constraints established by HIC experiments, labeled as "flow data 2003"\cite{danielewicz2002sci} and "FOPI 2016"~\cite{fevre2016npa}. BHF calculations yield EOSs that are too soft to satisfy the constraints imposed by HIC experiments. Conversely, RBHF calculations, attributed to significant relativistic repulsion, generate EOSs that better align with the experimental constraints.
    Moreover, the causality is automatically encoded in relativistic 
    calculations, manifested in $c_s^2 = \frac{\partial P}{\partial \epsilon}<1$,
    with $\epsilon$ being the energy density.

\section{SUMMARY AND OUTLOOK}\label{OBEPL-summary}
We constructed one-boson-exchange potentials by introducing exponential regulators with relative momentum cutoffs ranging from $\Lambda=\infty$ to 2 fm$^{-1}$. The regulator effectively suppress high momenta beyond the given cutoff $\Lambda$, so we name the potential with each $\Lambda$ ``OBEP$\Lambda$''. The parameters within the OBEP$\Lambda$s were fitted to $NN$ scattering phase shifts, low-energy scattering data, and deuteron binding energy. Notably, for $\Lambda\leqslant 3$ fm$^{-1}$, the potential matrix elements of our OBEP$\Lambda$s exhibit quantitative agreement with other realistic $NN$ potentials evolved using the renormalization group (RG) method.

Both NR-BHF and RBHF calculations were conducted using these OBEP$\Lambda$s. The equations of state obtained from BHF calculations at all cutoffs display an overbound nature. Notably, for $\Lambda\leqslant 3$ fm$^{-1}$, the corresponding  OBEP$\Lambda$ demonstrate softness that prevents the production of saturation phenomena, aligning with prior studies emphasizing the significance of the three-nucleon force (3NF) in achieving nuclear matter saturation. Conversely, all EOSs derived from RBHF calculations exhibit saturation behaviors; however, their saturation densities and binding energies slightly fall short of accurately reproducing empirical saturation properties. Further investigations will explore the contributions arising from the relativistic three-hole-line component in the BBG expansion and the impact of genuine relativistic 3NFs.

We examined the relativistic saturation mechanism by analyzing the kinetic and potential terms within relativistic definitions. As the Dirac mass impacts the lower component of the spinor, high-density relativistic kinetic terms exhibit negativity, while significant repulsion arises in the potential terms due to the interplay between the attractive scalar self-energy and the repulsive time-component vector self-energy. Remarkably, even exceedingly soft potentials can generate substantial scalar and time-component vector self-energies, reaching several hundreds of MeV. The suppression of the attractive scalar self-energy and the linear growth of the repulsive time-component vector self-energy collectively contribute to the stiffening of the relativistic EOS without the incorporation of 3NFs. This scenario could lead to divergent interpretations of nuclear matter saturation and related phenomena.
 


\bibliography{ref}

\begin{thebibliography}{61}%
\makeatletter
\providecommand \@ifxundefined [1]{%
 \@ifx{#1\undefined}
}%
\providecommand \@ifnum [1]{%
 \ifnum #1\expandafter \@firstoftwo
 \else \expandafter \@secondoftwo
 \fi
}%
\providecommand \@ifx [1]{%
 \ifx #1\expandafter \@firstoftwo
 \else \expandafter \@secondoftwo
 \fi
}%
\providecommand \natexlab [1]{#1}%
\providecommand \enquote  [1]{``#1''}%
\providecommand \bibnamefont  [1]{#1}%
\providecommand \bibfnamefont [1]{#1}%
\providecommand \citenamefont [1]{#1}%
\providecommand \href@noop [0]{\@secondoftwo}%
\providecommand \href [0]{\begingroup \@sanitize@url \@href}%
\providecommand \@href[1]{\@@startlink{#1}\@@href}%
\providecommand \@@href[1]{\endgroup#1\@@endlink}%
\providecommand \@sanitize@url [0]{\catcode `\\12\catcode `\$12\catcode
  `\&12\catcode `\#12\catcode `\^12\catcode `\_12\catcode `\%12\relax}%
\providecommand \@@startlink[1]{}%
\providecommand \@@endlink[0]{}%
\providecommand \url  [0]{\begingroup\@sanitize@url \@url }%
\providecommand \@url [1]{\endgroup\@href {#1}{\urlprefix }}%
\providecommand \urlprefix  [0]{URL }%
\providecommand \Eprint [0]{\href }%
\providecommand \doibase [0]{https://doi.org/}%
\providecommand \selectlanguage [0]{\@gobble}%
\providecommand \bibinfo  [0]{\@secondoftwo}%
\providecommand \bibfield  [0]{\@secondoftwo}%
\providecommand \translation [1]{[#1]}%
\providecommand \BibitemOpen [0]{}%
\providecommand \bibitemStop [0]{}%
\providecommand \bibitemNoStop [0]{.\EOS\space}%
\providecommand \EOS [0]{\spacefactor3000\relax}%
\providecommand \BibitemShut  [1]{\csname bibitem#1\endcsname}%
\let\auto@bib@innerbib\@empty
\bibitem [{\citenamefont {Erkelenz}(1974)}]{erkelenz1974pr}%
  \BibitemOpen
  \bibfield  {author} {\bibinfo {author} {\bibfnamefont {K.}~\bibnamefont
  {Erkelenz}},\ }\bibfield  {title} {\bibinfo {title} {Current status of the
  relativistic two-nucleon one boson exchange potential},\ }\href
  {https://doi.org/10.1016/0370-1573(74)90008-8} {\bibfield  {journal}
  {\bibinfo  {journal} {Phys. Rep.}\ }\textbf {\bibinfo {volume} {13}},\
  \bibinfo {pages} {191} (\bibinfo {year} {1974})}\BibitemShut {NoStop}%
\bibitem [{\citenamefont {Machleidt}\ \emph {et~al.}(1987)\citenamefont
  {Machleidt}, \citenamefont {Holinde},\ and\ \citenamefont
  {Elster}}]{machleidt1987pr}%
  \BibitemOpen
  \bibfield  {author} {\bibinfo {author} {\bibfnamefont {R.}~\bibnamefont
  {Machleidt}}, \bibinfo {author} {\bibfnamefont {K.}~\bibnamefont {Holinde}},\
  and\ \bibinfo {author} {\bibfnamefont {C.}~\bibnamefont {Elster}},\
  }\bibfield  {title} {\bibinfo {title} {The bonn meson-exchange model for the
  nucleon-nucleon interaction},\ }\href
  {https://doi.org/10.1016/S0370-1573(87)80002-9} {\bibfield  {journal}
  {\bibinfo  {journal} {Physics Reports}\ }\textbf {\bibinfo {volume} {149}},\
  \bibinfo {pages} {1} (\bibinfo {year} {1987})}\BibitemShut {NoStop}%
\bibitem [{\citenamefont {Machleidt}(1989)}]{machleidt1989anp}%
  \BibitemOpen
  \bibfield  {author} {\bibinfo {author} {\bibfnamefont {R.}~\bibnamefont
  {Machleidt}},\ }\bibinfo {title} {The meson theory of nuclear forces and
  nuclear structure},\ in\ \href {https://doi.org/10.1007/978-1-4613-9907-0_2}
  {\emph {\bibinfo {booktitle} {Adv. Nucl. Phys.}}},\ Vol.~\bibinfo {volume}
  {19},\ \bibinfo {editor} {edited by\ \bibinfo {editor} {\bibfnamefont
  {e.~a.}\ \bibnamefont {J.~W.~Negele}}}\ (\bibinfo  {publisher} {Springer},\
  \bibinfo {address} {USA},\ \bibinfo {year} {1989})\ pp.\ \bibinfo {pages}
  {189--376}\BibitemShut {NoStop}%
\bibitem [{\citenamefont {Brockmann}\ and\ \citenamefont
  {Machleidt}(1990)}]{brockmann1990prc}%
  \BibitemOpen
  \bibfield  {author} {\bibinfo {author} {\bibfnamefont {R.}~\bibnamefont
  {Brockmann}}\ and\ \bibinfo {author} {\bibfnamefont {R.}~\bibnamefont
  {Machleidt}},\ }\bibfield  {title} {\bibinfo {title} {Relativistic nuclear
  structure. i. nuclear matter},\ }\href
  {https://doi.org/10.1103/PhysRevC.42.1965} {\bibfield  {journal} {\bibinfo
  {journal} {Phys. Rev. C}\ }\textbf {\bibinfo {volume} {42}},\ \bibinfo
  {pages} {1965} (\bibinfo {year} {1990})}\BibitemShut {NoStop}%
\bibitem [{\citenamefont {Machleidt}(2001)}]{machleidt2001prc}%
  \BibitemOpen
  \bibfield  {author} {\bibinfo {author} {\bibfnamefont {R.}~\bibnamefont
  {Machleidt}},\ }\bibfield  {title} {\bibinfo {title} {High-precision,
  charge-dependent bonn nucleon-nucleon potential},\ }\href
  {https://doi.org/10.1103/PhysRevC.63.024001} {\bibfield  {journal} {\bibinfo
  {journal} {Physical Review C}\ }\textbf {\bibinfo {volume} {63}},\ \bibinfo
  {pages} {024001} (\bibinfo {year} {2001})}\BibitemShut {NoStop}%
\bibitem [{\citenamefont {Gross-Boelting}\ \emph {et~al.}(1999)\citenamefont
  {Gross-Boelting}, \citenamefont {Fuchs},\ and\ \citenamefont
  {Faessler}}]{gross-beolting1999npa}%
  \BibitemOpen
  \bibfield  {author} {\bibinfo {author} {\bibfnamefont {T.}~\bibnamefont
  {Gross-Boelting}}, \bibinfo {author} {\bibfnamefont {C.}~\bibnamefont
  {Fuchs}},\ and\ \bibinfo {author} {\bibfnamefont {A.}~\bibnamefont
  {Faessler}},\ }\bibfield  {title} {\bibinfo {title} {Covariant representation
  of the relativistic brueckner t-matrix and nuclear matter problem},\ }\href
  {https://doi.org/10.106/S0375-9474(99)00022-6} {\bibfield  {journal}
  {\bibinfo  {journal} {Nuclear Physics A}\ }\textbf {\bibinfo {volume}
  {648}},\ \bibinfo {pages} {105} (\bibinfo {year} {1999})}\BibitemShut
  {NoStop}%
\bibitem [{\citenamefont {Wang}\ \emph
  {et~al.}(2020{\natexlab{a}})\citenamefont {Wang}, \citenamefont {Hu},
  \citenamefont {Zhang},\ and\ \citenamefont {Shen}}]{wang2020cpc}%
  \BibitemOpen
  \bibfield  {author} {\bibinfo {author} {\bibfnamefont {C.~C.}\ \bibnamefont
  {Wang}}, \bibinfo {author} {\bibfnamefont {J.~N.}\ \bibnamefont {Hu}},
  \bibinfo {author} {\bibfnamefont {Y.}~\bibnamefont {Zhang}},\ and\ \bibinfo
  {author} {\bibfnamefont {H.}~\bibnamefont {Shen}},\ }\bibfield  {title}
  {\bibinfo {title} {The charge-dependent bonn potentials with pseudovector
  pion-nucleon coupling},\ }\href
  {https://doi.org/10.1088/1674-1137/43/11/114107} {\bibfield  {journal}
  {\bibinfo  {journal} {Chin. Phys. C}\ }\textbf {\bibinfo {volume} {43}},\
  \bibinfo {pages} {114107} (\bibinfo {year} {2020}{\natexlab{a}})}\BibitemShut
  {NoStop}%
\bibitem [{\citenamefont {Wang}\ \emph
  {et~al.}(2020{\natexlab{b}})\citenamefont {Wang}, \citenamefont {Hu},
  \citenamefont {Zhang},\ and\ \citenamefont {Shen}}]{wang2020jpg}%
  \BibitemOpen
  \bibfield  {author} {\bibinfo {author} {\bibfnamefont {C.~C.}\ \bibnamefont
  {Wang}}, \bibinfo {author} {\bibfnamefont {J.~N.}\ \bibnamefont {Hu}},
  \bibinfo {author} {\bibfnamefont {Y.}~\bibnamefont {Zhang}},\ and\ \bibinfo
  {author} {\bibfnamefont {H.}~\bibnamefont {Shen}},\ }\bibfield  {title}
  {\bibinfo {title} {Properties of nuclear matter in relativistic
  brueckner–hartree–fock model with high-precision charge-dependent
  potentials},\ }\href {https://doi.org/10.1088/1361-6471/aba423} {\bibfield
  {journal} {\bibinfo  {journal} {J. Phys. G}\ }\textbf {\bibinfo {volume}
  {47}},\ \bibinfo {pages} {105108} (\bibinfo {year}
  {2020}{\natexlab{b}})}\BibitemShut {NoStop}%
\bibitem [{\citenamefont {Epelbaum}\ \emph {et~al.}(1998)\citenamefont
  {Epelbaum}, \citenamefont {Glöckle},\ and\ \citenamefont
  {Meißner}}]{epelbaum1998plb}%
  \BibitemOpen
  \bibfield  {author} {\bibinfo {author} {\bibfnamefont {E.}~\bibnamefont
  {Epelbaum}}, \bibinfo {author} {\bibfnamefont {W.}~\bibnamefont {Glöckle}},\
  and\ \bibinfo {author} {\bibfnamefont {U.-G.}\ \bibnamefont {Meißner}},\
  }\bibfield  {title} {\bibinfo {title} {Low-momentum effective theory for
  nucleons},\ }\href
  {https://doi.org/https://doi.org/10.1016/S0370-2693(98)01045-4} {\bibfield
  {journal} {\bibinfo  {journal} {Phys. Lett. B}\ }\textbf {\bibinfo {volume}
  {439}},\ \bibinfo {pages} {1} (\bibinfo {year} {1998})}\BibitemShut {NoStop}%
\bibitem [{\citenamefont {Epelbaum}\ \emph {et~al.}(1999)\citenamefont
  {Epelbaum}, \citenamefont {Glöckle}, \citenamefont {Krüger},\ and\
  \citenamefont {Meißner}}]{epelbaum1999npa}%
  \BibitemOpen
  \bibfield  {author} {\bibinfo {author} {\bibfnamefont {E.}~\bibnamefont
  {Epelbaum}}, \bibinfo {author} {\bibfnamefont {W.}~\bibnamefont {Glöckle}},
  \bibinfo {author} {\bibfnamefont {A.}~\bibnamefont {Krüger}},\ and\ \bibinfo
  {author} {\bibfnamefont {U.-G.}\ \bibnamefont {Meißner}},\ }\bibfield
  {title} {\bibinfo {title} {Effective theory for the two-nucleon system},\
  }\href {https://doi.org/10.1016/S0375-9474(98)00585-5} {\bibfield  {journal}
  {\bibinfo  {journal} {Nucl. Phys. A}\ }\textbf {\bibinfo {volume} {645}},\
  \bibinfo {pages} {413} (\bibinfo {year} {1999})}\BibitemShut {NoStop}%
\bibitem [{\citenamefont {Epelbaum}\ \emph {et~al.}(2009)\citenamefont
  {Epelbaum}, \citenamefont {Hammer},\ and\ \citenamefont
  {Mei\ss{}ner}}]{epelbaum2009rmp}%
  \BibitemOpen
  \bibfield  {author} {\bibinfo {author} {\bibfnamefont {E.}~\bibnamefont
  {Epelbaum}}, \bibinfo {author} {\bibfnamefont {H.-W.}\ \bibnamefont
  {Hammer}},\ and\ \bibinfo {author} {\bibfnamefont {U.-G.}\ \bibnamefont
  {Mei\ss{}ner}},\ }\bibfield  {title} {\bibinfo {title} {Modern theory of
  nuclear forces},\ }\href {https://doi.org/10.1103/RevModPhys.81.1773}
  {\bibfield  {journal} {\bibinfo  {journal} {Rev. Mod. Phys.}\ }\textbf
  {\bibinfo {volume} {81}},\ \bibinfo {pages} {1773} (\bibinfo {year}
  {2009})}\BibitemShut {NoStop}%
\bibitem [{\citenamefont {Bogner}\ \emph {et~al.}(2010)\citenamefont {Bogner},
  \citenamefont {Furnstahl},\ and\ \citenamefont {Schwenk}}]{bogner2010ppnp}%
  \BibitemOpen
  \bibfield  {author} {\bibinfo {author} {\bibfnamefont {S.~K.}\ \bibnamefont
  {Bogner}}, \bibinfo {author} {\bibfnamefont {R.~J.}\ \bibnamefont
  {Furnstahl}},\ and\ \bibinfo {author} {\bibfnamefont {A.}~\bibnamefont
  {Schwenk}},\ }\bibfield  {title} {\bibinfo {title} {From low-momentum
  interactions to nuclear structure},\ }\href
  {https://doi.org/10.1016/j.ppnp.2010.03.001} {\bibfield  {journal} {\bibinfo
  {journal} {Prog. Part. Nucl. Phys.}\ }\textbf {\bibinfo {volume} {65}},\
  \bibinfo {pages} {94} (\bibinfo {year} {2010})}\BibitemShut {NoStop}%
\bibitem [{\citenamefont {Machleidt}\ and\ \citenamefont
  {Entem}(2011)}]{machleidt2011pr}%
  \BibitemOpen
  \bibfield  {author} {\bibinfo {author} {\bibfnamefont {R.}~\bibnamefont
  {Machleidt}}\ and\ \bibinfo {author} {\bibfnamefont {D.~R.}\ \bibnamefont
  {Entem}},\ }\bibfield  {title} {\bibinfo {title} {Chiral effective field
  theory and nuclear forces},\ }\href
  {https://doi.org/10.1016/j.physrep.2011.02.001} {\bibfield  {journal}
  {\bibinfo  {journal} {Phys. Rep.}\ }\textbf {\bibinfo {volume} {503}},\
  \bibinfo {pages} {1} (\bibinfo {year} {2011})}\BibitemShut {NoStop}%
\bibitem [{\citenamefont {Epelbaum}\ \emph {et~al.}(2005)\citenamefont
  {Epelbaum}, \citenamefont {Glöckle},\ and\ \citenamefont
  {Meißner}}]{epelbaum2005npa}%
  \BibitemOpen
  \bibfield  {author} {\bibinfo {author} {\bibfnamefont {E.}~\bibnamefont
  {Epelbaum}}, \bibinfo {author} {\bibfnamefont {W.}~\bibnamefont {Glöckle}},\
  and\ \bibinfo {author} {\bibfnamefont {U.-G.}\ \bibnamefont {Meißner}},\
  }\bibfield  {title} {\bibinfo {title} {The two-nucleon system at
  next-to-next-to-next-to-leading order},\ }\href
  {https://doi.org/10.1016/j.nuclphysa.2004.09.107} {\bibfield  {journal}
  {\bibinfo  {journal} {Nucl. Phys. A}\ }\textbf {\bibinfo {volume} {747}},\
  \bibinfo {pages} {362} (\bibinfo {year} {2005})}\BibitemShut {NoStop}%
\bibitem [{\citenamefont {Hebeler}\ \emph {et~al.}(2011)\citenamefont
  {Hebeler}, \citenamefont {Bogner}, \citenamefont {Furnstahl}, \citenamefont
  {Nogga},\ and\ \citenamefont {Schwenk}}]{hebeler2011prc}%
  \BibitemOpen
  \bibfield  {author} {\bibinfo {author} {\bibfnamefont {K.}~\bibnamefont
  {Hebeler}}, \bibinfo {author} {\bibfnamefont {S.~K.}\ \bibnamefont {Bogner}},
  \bibinfo {author} {\bibfnamefont {R.~J.}\ \bibnamefont {Furnstahl}}, \bibinfo
  {author} {\bibfnamefont {A.}~\bibnamefont {Nogga}},\ and\ \bibinfo {author}
  {\bibfnamefont {A.}~\bibnamefont {Schwenk}},\ }\bibfield  {title} {\bibinfo
  {title} {Improved nuclear matter calculations from chiral low-momentum
  interactions},\ }\href {https://doi.org/10.1103/PhysRevC.83.031301}
  {\bibfield  {journal} {\bibinfo  {journal} {Phys. Rev. C}\ }\textbf {\bibinfo
  {volume} {83}},\ \bibinfo {pages} {031301} (\bibinfo {year}
  {2011})}\BibitemShut {NoStop}%
\bibitem [{\citenamefont {Epelbaum}(2006)}]{epelabum2006ppnp}%
  \BibitemOpen
  \bibfield  {author} {\bibinfo {author} {\bibfnamefont {E.}~\bibnamefont
  {Epelbaum}},\ }\bibfield  {title} {\bibinfo {title} {Few-nucleon forces and
  systems in chiral effective field theory},\ }\href
  {https://doi.org/10.1016/j.ppnp.2005.09.002} {\bibfield  {journal} {\bibinfo
  {journal} {Prog. Part. Nucl. Phys.}\ }\textbf {\bibinfo {volume} {57}},\
  \bibinfo {pages} {654} (\bibinfo {year} {2006})}\BibitemShut {NoStop}%
\bibitem [{\citenamefont {Bogner}\ \emph
  {et~al.}(2008{\natexlab{a}})\citenamefont {Bogner}, \citenamefont
  {Furnstahl},\ and\ \citenamefont {Perry}}]{bogner2008annphys}%
  \BibitemOpen
  \bibfield  {author} {\bibinfo {author} {\bibfnamefont {S.~K.}\ \bibnamefont
  {Bogner}}, \bibinfo {author} {\bibfnamefont {R.~J.}\ \bibnamefont
  {Furnstahl}},\ and\ \bibinfo {author} {\bibfnamefont {R.~J.}\ \bibnamefont
  {Perry}},\ }\bibfield  {title} {\bibinfo {title} {Three-body forces produced
  by a similarity renormalization group transformation in a simple model},\
  }\href {https://doi.org/10.1016/j.aop.2007.09.001} {\bibfield  {journal}
  {\bibinfo  {journal} {Ann. Phys.}\ }\textbf {\bibinfo {volume} {323}},\
  \bibinfo {pages} {1478} (\bibinfo {year} {2008}{\natexlab{a}})}\BibitemShut
  {NoStop}%
\bibitem [{\citenamefont {Hebeler}(2021)}]{hebeler2021pr}%
  \BibitemOpen
  \bibfield  {author} {\bibinfo {author} {\bibfnamefont {K.}~\bibnamefont
  {Hebeler}},\ }\bibfield  {title} {\bibinfo {title} {Three-nucleon forces:
  Implementation and applications to atomic nuclei and dense matter},\ }\href
  {https://doi.org/https://doi.org/10.1016/j.physrep.2020.08.009} {\bibfield
  {journal} {\bibinfo  {journal} {Phys. Rep.}\ }\textbf {\bibinfo {volume}
  {890}},\ \bibinfo {pages} {1} (\bibinfo {year} {2021})}\BibitemShut {NoStop}%
\bibitem [{\citenamefont {Nogga}\ \emph {et~al.}(2004)\citenamefont {Nogga},
  \citenamefont {Bogner},\ and\ \citenamefont {Schwenk}}]{noggaprc2004}%
  \BibitemOpen
  \bibfield  {author} {\bibinfo {author} {\bibfnamefont {A.}~\bibnamefont
  {Nogga}}, \bibinfo {author} {\bibfnamefont {S.~K.}\ \bibnamefont {Bogner}},\
  and\ \bibinfo {author} {\bibfnamefont {A.}~\bibnamefont {Schwenk}},\
  }\bibfield  {title} {\bibinfo {title} {Low-momentum interaction in
  few-nucleon systems},\ }\href {https://doi.org/10.1103/PhysRevC.70.061002}
  {\bibfield  {journal} {\bibinfo  {journal} {Phys. Rev. C}\ }\textbf {\bibinfo
  {volume} {70}},\ \bibinfo {pages} {061002} (\bibinfo {year}
  {2004})}\BibitemShut {NoStop}%
\bibitem [{\citenamefont {Bogner}\ \emph
  {et~al.}(2008{\natexlab{b}})\citenamefont {Bogner}, \citenamefont
  {Furnstahl}, \citenamefont {Maris}, \citenamefont {Perry}, \citenamefont
  {Schwenk},\ and\ \citenamefont {Vary}}]{bogner2008npa}%
  \BibitemOpen
  \bibfield  {author} {\bibinfo {author} {\bibfnamefont {S.~K.}\ \bibnamefont
  {Bogner}}, \bibinfo {author} {\bibfnamefont {R.~J.}\ \bibnamefont
  {Furnstahl}}, \bibinfo {author} {\bibfnamefont {P.}~\bibnamefont {Maris}},
  \bibinfo {author} {\bibfnamefont {R.~J.}\ \bibnamefont {Perry}}, \bibinfo
  {author} {\bibfnamefont {A.}~\bibnamefont {Schwenk}},\ and\ \bibinfo {author}
  {\bibfnamefont {J.~P.}\ \bibnamefont {Vary}},\ }\bibfield  {title} {\bibinfo
  {title} {Convergence in the no-core shell model with low-momentum two-nucleon
  interactions},\ }\href {https://doi.org/10.1016/j.nuclphysa.2007.12.008}
  {\bibfield  {journal} {\bibinfo  {journal} {Nucl. Phys. A}\ }\textbf
  {\bibinfo {volume} {801}},\ \bibinfo {pages} {21} (\bibinfo {year}
  {2008}{\natexlab{b}})}\BibitemShut {NoStop}%
\bibitem [{\citenamefont {Tews}\ \emph {et~al.}(2013)\citenamefont {Tews},
  \citenamefont {Kr\"uger}, \citenamefont {Hebeler},\ and\ \citenamefont
  {Schwenk}}]{tew2013prl}%
  \BibitemOpen
  \bibfield  {author} {\bibinfo {author} {\bibfnamefont {I.}~\bibnamefont
  {Tews}}, \bibinfo {author} {\bibfnamefont {T.}~\bibnamefont {Kr\"uger}},
  \bibinfo {author} {\bibfnamefont {K.}~\bibnamefont {Hebeler}},\ and\ \bibinfo
  {author} {\bibfnamefont {A.}~\bibnamefont {Schwenk}},\ }\bibfield  {title}
  {\bibinfo {title} {Neutron matter at next-to-next-to-next-to-leading order in
  chiral effective field theory},\ }\href
  {https://doi.org/10.1103/PhysRevLett.110.032504} {\bibfield  {journal}
  {\bibinfo  {journal} {Phys. Rev. Lett.}\ }\textbf {\bibinfo {volume} {110}},\
  \bibinfo {pages} {032504} (\bibinfo {year} {2013})}\BibitemShut {NoStop}%
\bibitem [{\citenamefont {Drischler}\ \emph {et~al.}(2016)\citenamefont
  {Drischler}, \citenamefont {Hebeler},\ and\ \citenamefont
  {Schwenk}}]{drischler2016prc}%
  \BibitemOpen
  \bibfield  {author} {\bibinfo {author} {\bibfnamefont {C.}~\bibnamefont
  {Drischler}}, \bibinfo {author} {\bibfnamefont {K.}~\bibnamefont {Hebeler}},\
  and\ \bibinfo {author} {\bibfnamefont {A.}~\bibnamefont {Schwenk}},\
  }\bibfield  {title} {\bibinfo {title} {Asymmetric nuclear matter based on
  chiral two- and three-nucleon interactions},\ }\href
  {https://doi.org/10.1103/PhysRevC.93.054314} {\bibfield  {journal} {\bibinfo
  {journal} {Phys. Rev. C}\ }\textbf {\bibinfo {volume} {93}},\ \bibinfo
  {pages} {054314} (\bibinfo {year} {2016})}\BibitemShut {NoStop}%
\bibitem [{\citenamefont {Drischler}\ \emph {et~al.}(2019)\citenamefont
  {Drischler}, \citenamefont {Hebeler},\ and\ \citenamefont
  {Schwenk}}]{drischler2019prl}%
  \BibitemOpen
  \bibfield  {author} {\bibinfo {author} {\bibfnamefont {C.}~\bibnamefont
  {Drischler}}, \bibinfo {author} {\bibfnamefont {K.}~\bibnamefont {Hebeler}},\
  and\ \bibinfo {author} {\bibfnamefont {A.}~\bibnamefont {Schwenk}},\
  }\bibfield  {title} {\bibinfo {title} {Chiral interactions up to
  next-to-next-to-next-to-leading order and nuclear saturation},\ }\href
  {https://doi.org/10.1103/PhysRevLett.122.042501} {\bibfield  {journal}
  {\bibinfo  {journal} {Phys. Rev. Lett.}\ }\textbf {\bibinfo {volume} {122}},\
  \bibinfo {pages} {042501} (\bibinfo {year} {2019})}\BibitemShut {NoStop}%
\bibitem [{\citenamefont {Simonis}\ \emph {et~al.}(2017)\citenamefont
  {Simonis}, \citenamefont {Stroberg}, \citenamefont {Hebeler}, \citenamefont
  {Holt},\ and\ \citenamefont {Schwenk}}]{simonis2017prc}%
  \BibitemOpen
  \bibfield  {author} {\bibinfo {author} {\bibfnamefont {J.}~\bibnamefont
  {Simonis}}, \bibinfo {author} {\bibfnamefont {S.~R.}\ \bibnamefont
  {Stroberg}}, \bibinfo {author} {\bibfnamefont {K.}~\bibnamefont {Hebeler}},
  \bibinfo {author} {\bibfnamefont {J.~D.}\ \bibnamefont {Holt}},\ and\
  \bibinfo {author} {\bibfnamefont {A.}~\bibnamefont {Schwenk}},\ }\bibfield
  {title} {\bibinfo {title} {Saturation with chiral interactions and
  consequences for finite nuclei},\ }\href
  {https://doi.org/10.1103/PhysRevC.96.014303} {\bibfield  {journal} {\bibinfo
  {journal} {Phys. Rev. C}\ }\textbf {\bibinfo {volume} {96}},\ \bibinfo
  {pages} {014303} (\bibinfo {year} {2017})}\BibitemShut {NoStop}%
\bibitem [{\citenamefont {Ekstr\"om}\ \emph {et~al.}(2018)\citenamefont
  {Ekstr\"om}, \citenamefont {Hagen}, \citenamefont {Morris}, \citenamefont
  {Papenbrock},\ and\ \citenamefont {Schwartz}}]{ekstroem2018prc}%
  \BibitemOpen
  \bibfield  {author} {\bibinfo {author} {\bibfnamefont {A.}~\bibnamefont
  {Ekstr\"om}}, \bibinfo {author} {\bibfnamefont {G.}~\bibnamefont {Hagen}},
  \bibinfo {author} {\bibfnamefont {T.~D.}\ \bibnamefont {Morris}}, \bibinfo
  {author} {\bibfnamefont {T.}~\bibnamefont {Papenbrock}},\ and\ \bibinfo
  {author} {\bibfnamefont {P.~D.}\ \bibnamefont {Schwartz}},\ }\bibfield
  {title} {\bibinfo {title} {$\mathrm{\ensuremath{\Delta}}$ isobars and nuclear
  saturation},\ }\href {https://doi.org/10.1103/PhysRevC.97.024332} {\bibfield
  {journal} {\bibinfo  {journal} {Phys. Rev. C}\ }\textbf {\bibinfo {volume}
  {97}},\ \bibinfo {pages} {024332} (\bibinfo {year} {2018})}\BibitemShut
  {NoStop}%
\bibitem [{\citenamefont {Hoppe}\ \emph {et~al.}(2019)\citenamefont {Hoppe},
  \citenamefont {Drischler}, \citenamefont {Hebeler}, \citenamefont {Schwenk},\
  and\ \citenamefont {Simonis}}]{hoppe2019prc}%
  \BibitemOpen
  \bibfield  {author} {\bibinfo {author} {\bibfnamefont {J.}~\bibnamefont
  {Hoppe}}, \bibinfo {author} {\bibfnamefont {C.}~\bibnamefont {Drischler}},
  \bibinfo {author} {\bibfnamefont {K.}~\bibnamefont {Hebeler}}, \bibinfo
  {author} {\bibfnamefont {A.}~\bibnamefont {Schwenk}},\ and\ \bibinfo {author}
  {\bibfnamefont {J.}~\bibnamefont {Simonis}},\ }\bibfield  {title} {\bibinfo
  {title} {Probing chiral interactions up to next-to-next-to-next-to-leading
  order in medium-mass nuclei},\ }\href
  {https://doi.org/10.1103/PhysRevC.100.024318} {\bibfield  {journal} {\bibinfo
   {journal} {Phys. Rev. C}\ }\textbf {\bibinfo {volume} {100}},\ \bibinfo
  {pages} {024318} (\bibinfo {year} {2019})}\BibitemShut {NoStop}%
\bibitem [{\citenamefont {Sammarruca}\ and\ \citenamefont
  {Millerson}(2020)}]{sammarruca2020prc}%
  \BibitemOpen
  \bibfield  {author} {\bibinfo {author} {\bibfnamefont {F.}~\bibnamefont
  {Sammarruca}}\ and\ \bibinfo {author} {\bibfnamefont {R.}~\bibnamefont
  {Millerson}},\ }\bibfield  {title} {\bibinfo {title} {Exploring the
  relationship between nuclear matter and finite nuclei with chiral two- and
  three-nucleon forces},\ }\href {https://doi.org/10.1103/PhysRevC.102.034313}
  {\bibfield  {journal} {\bibinfo  {journal} {Phys. Rev. C}\ }\textbf {\bibinfo
  {volume} {102}},\ \bibinfo {pages} {034313} (\bibinfo {year}
  {2020})}\BibitemShut {NoStop}%
\bibitem [{\citenamefont {Lattimer}(2012)}]{lattimer2012arnps}%
  \BibitemOpen
  \bibfield  {author} {\bibinfo {author} {\bibfnamefont {J.~M.}\ \bibnamefont
  {Lattimer}},\ }\bibfield  {title} {\bibinfo {title} {The nuclear equation of
  state and neutron star masses},\ }\href
  {https://doi.org/10.1146/annurev-nucl-102711-095018} {\bibfield  {journal}
  {\bibinfo  {journal} {Annu. Rev. Nucl. Part. Sci.}\ }\textbf {\bibinfo
  {volume} {62}},\ \bibinfo {pages} {485} (\bibinfo {year} {2012})}\BibitemShut
  {NoStop}%
\bibitem [{\citenamefont {Oertel}\ \emph {et~al.}(2017)\citenamefont {Oertel},
  \citenamefont {Hempel}, \citenamefont {Kl\"ahn},\ and\ \citenamefont
  {Typel}}]{oertel2017rmp}%
  \BibitemOpen
  \bibfield  {author} {\bibinfo {author} {\bibfnamefont {M.}~\bibnamefont
  {Oertel}}, \bibinfo {author} {\bibfnamefont {M.}~\bibnamefont {Hempel}},
  \bibinfo {author} {\bibfnamefont {T.}~\bibnamefont {Kl\"ahn}},\ and\ \bibinfo
  {author} {\bibfnamefont {S.}~\bibnamefont {Typel}},\ }\bibfield  {title}
  {\bibinfo {title} {Equations of state for supernovae and compact stars},\
  }\href {https://doi.org/10.1103/RevModPhys.89.015007} {\bibfield  {journal}
  {\bibinfo  {journal} {Rev. Mod. Phys.}\ }\textbf {\bibinfo {volume} {89}},\
  \bibinfo {pages} {015007} (\bibinfo {year} {2017})}\BibitemShut {NoStop}%
\bibitem [{\citenamefont {Carreau}\ \emph {et~al.}(2019)\citenamefont
  {Carreau}, \citenamefont {Gulminelli},\ and\ \citenamefont
  {Margueron}}]{carreau2019prc}%
  \BibitemOpen
  \bibfield  {author} {\bibinfo {author} {\bibfnamefont {T.}~\bibnamefont
  {Carreau}}, \bibinfo {author} {\bibfnamefont {F.}~\bibnamefont
  {Gulminelli}},\ and\ \bibinfo {author} {\bibfnamefont {J.}~\bibnamefont
  {Margueron}},\ }\bibfield  {title} {\bibinfo {title} {General predictions for
  the neutron star crustal moment of inertia},\ }\href
  {https://doi.org/10.1103/PhysRevC.100.055803} {\bibfield  {journal} {\bibinfo
   {journal} {Phys. Rev. C}\ }\textbf {\bibinfo {volume} {100}},\ \bibinfo
  {pages} {055803} (\bibinfo {year} {2019})}\BibitemShut {NoStop}%
\bibitem [{\citenamefont {Wang}\ \emph
  {et~al.}(2020{\natexlab{c}})\citenamefont {Wang}, \citenamefont {Hu},
  \citenamefont {Zhang},\ and\ \citenamefont {Shen}}]{wang2020apj}%
  \BibitemOpen
  \bibfield  {author} {\bibinfo {author} {\bibfnamefont {C.~C.}\ \bibnamefont
  {Wang}}, \bibinfo {author} {\bibfnamefont {J.}~\bibnamefont {Hu}}, \bibinfo
  {author} {\bibfnamefont {Y.}~\bibnamefont {Zhang}},\ and\ \bibinfo {author}
  {\bibfnamefont {H.}~\bibnamefont {Shen}},\ }\bibfield  {title} {\bibinfo
  {title} {Properties of neutron stars described by a relativistic ab initio
  model},\ }\href {https://doi.org/10.3847/1538-4357/ab994b} {\bibfield
  {journal} {\bibinfo  {journal} {Astrophys. J.}\ }\textbf {\bibinfo {volume}
  {897}},\ \bibinfo {pages} {96} (\bibinfo {year}
  {2020}{\natexlab{c}})}\BibitemShut {NoStop}%
\bibitem [{\citenamefont {Danielewicz}\ \emph {et~al.}(2002)\citenamefont
  {Danielewicz}, \citenamefont {Lacey},\ and\ \citenamefont
  {Lynch}}]{danielewicz2002sci}%
  \BibitemOpen
  \bibfield  {author} {\bibinfo {author} {\bibfnamefont {P.}~\bibnamefont
  {Danielewicz}}, \bibinfo {author} {\bibfnamefont {R.}~\bibnamefont {Lacey}},\
  and\ \bibinfo {author} {\bibfnamefont {W.~G.}\ \bibnamefont {Lynch}},\
  }\bibfield  {title} {\bibinfo {title} {Determination of the equation of state
  of dense matter},\ }\href {https://doi.org/10.1126/science.1078070}
  {\bibfield  {journal} {\bibinfo  {journal} {Sci}\ }\textbf {\bibinfo {volume}
  {298}},\ \bibinfo {pages} {1592} (\bibinfo {year} {2002})}\BibitemShut
  {NoStop}%
\bibitem [{\citenamefont {Le~Fèvre}\ \emph {et~al.}(2016)\citenamefont
  {Le~Fèvre}, \citenamefont {Leifels}, \citenamefont {Reisdorf}, \citenamefont
  {Aichelin},\ and\ \citenamefont {Hartnack}}]{fevre2016npa}%
  \BibitemOpen
  \bibfield  {author} {\bibinfo {author} {\bibfnamefont {A.}~\bibnamefont
  {Le~Fèvre}}, \bibinfo {author} {\bibfnamefont {Y.}~\bibnamefont {Leifels}},
  \bibinfo {author} {\bibfnamefont {W.}~\bibnamefont {Reisdorf}}, \bibinfo
  {author} {\bibfnamefont {J.}~\bibnamefont {Aichelin}},\ and\ \bibinfo
  {author} {\bibfnamefont {C.}~\bibnamefont {Hartnack}},\ }\bibfield  {title}
  {\bibinfo {title} {Constraining the nuclear matter equation of state around
  twice saturation density},\ }\href
  {https://doi.org/10.1016/j.nuclphysa.2015.09.015} {\bibfield  {journal}
  {\bibinfo  {journal} {Nucl. Phys. A}\ }\textbf {\bibinfo {volume} {945}},\
  \bibinfo {pages} {112} (\bibinfo {year} {2016})}\BibitemShut {NoStop}%
\bibitem [{\citenamefont {Wolter}\ \emph {et~al.}(2022)\citenamefont {Wolter}
  \emph {et~al.}}]{hermann2022ppnp}%
  \BibitemOpen
  \bibfield  {author} {\bibinfo {author} {\bibfnamefont {H.}~\bibnamefont
  {Wolter}} \emph {et~al.} (\bibinfo {collaboration} {TMEP}),\ }\bibfield
  {title} {\bibinfo {title} {{Transport model comparison studies of
  intermediate-energy heavy-ion collisions}},\ }\href
  {https://doi.org/10.1016/j.ppnp.2022.103962} {\bibfield  {journal} {\bibinfo
  {journal} {Prog. Part. Nucl. Phys.}\ }\textbf {\bibinfo {volume} {125}},\
  \bibinfo {pages} {103962} (\bibinfo {year} {2022})}\BibitemShut {NoStop}%
\bibitem [{\citenamefont {Day}(1978)}]{day1978rmp}%
  \BibitemOpen
  \bibfield  {author} {\bibinfo {author} {\bibfnamefont {B.~D.}\ \bibnamefont
  {Day}},\ }\bibfield  {title} {\bibinfo {title} {Current state of nuclear
  matter calculations},\ }\href {https://doi.org/10.1103/RevModPhys.50.495}
  {\bibfield  {journal} {\bibinfo  {journal} {Rev. Mod. Phys.}\ }\textbf
  {\bibinfo {volume} {50}},\ \bibinfo {pages} {495} (\bibinfo {year}
  {1978})}\BibitemShut {NoStop}%
\bibitem [{\citenamefont {Pandharipande}\ and\ \citenamefont
  {Wiringa}(1979)}]{pandharipande1979rmp}%
  \BibitemOpen
  \bibfield  {author} {\bibinfo {author} {\bibfnamefont {V.~R.}\ \bibnamefont
  {Pandharipande}}\ and\ \bibinfo {author} {\bibfnamefont {R.~B.}\ \bibnamefont
  {Wiringa}},\ }\bibfield  {title} {\bibinfo {title} {Variations on a theme of
  nuclear matter},\ }\href {https://doi.org/10.1103/RevModPhys.51.821}
  {\bibfield  {journal} {\bibinfo  {journal} {Rev. Mod. Phys.}\ }\textbf
  {\bibinfo {volume} {51}},\ \bibinfo {pages} {821} (\bibinfo {year}
  {1979})}\BibitemShut {NoStop}%
\bibitem [{\citenamefont {Clark}(1979)}]{clark1979ppnp}%
  \BibitemOpen
  \bibfield  {author} {\bibinfo {author} {\bibfnamefont {J.~W.}\ \bibnamefont
  {Clark}},\ }\bibfield  {title} {\bibinfo {title} {Variational theory of
  nuclear matter},\ }\href
  {https://doi.org/https://doi.org/10.1016/0146-6410(79)90004-8} {\bibfield
  {journal} {\bibinfo  {journal} {Prog. Part. Nucl. Phys.}\ }\textbf {\bibinfo
  {volume} {2}},\ \bibinfo {pages} {89} (\bibinfo {year} {1979})}\BibitemShut
  {NoStop}%
\bibitem [{\citenamefont {Day}(1979)}]{day1979npa}%
  \BibitemOpen
  \bibfield  {author} {\bibinfo {author} {\bibfnamefont {B.~D.}\ \bibnamefont
  {Day}},\ }\bibfield  {title} {\bibinfo {title} {Brueckner-bethe calculations
  of nuclear matter},\ }\href {https://doi.org/10.1016/0375-9474(79)90208-2}
  {\bibfield  {journal} {\bibinfo  {journal} {Nucl. Phys. A}\ }\textbf
  {\bibinfo {volume} {328}},\ \bibinfo {pages} {1} (\bibinfo {year}
  {1979})}\BibitemShut {NoStop}%
\bibitem [{\citenamefont {Hjorth-Jensen}\ \emph {et~al.}(1995)\citenamefont
  {Hjorth-Jensen}, \citenamefont {Kuo},\ and\ \citenamefont
  {Osnes}}]{jensen1995pr}%
  \BibitemOpen
  \bibfield  {author} {\bibinfo {author} {\bibfnamefont {M.}~\bibnamefont
  {Hjorth-Jensen}}, \bibinfo {author} {\bibfnamefont {T.~T.~S.}\ \bibnamefont
  {Kuo}},\ and\ \bibinfo {author} {\bibfnamefont {E.}~\bibnamefont {Osnes}},\
  }\bibfield  {title} {\bibinfo {title} {Realistic effective interactions for
  nuclear systems},\ }\href {https://doi.org/10.1016/0370-1573(95)00012-6}
  {\bibfield  {journal} {\bibinfo  {journal} {Phys. Rep.}\ }\textbf {\bibinfo
  {volume} {261}},\ \bibinfo {pages} {125} (\bibinfo {year}
  {1995})}\BibitemShut {NoStop}%
\bibitem [{\citenamefont {Coester}\ \emph {et~al.}(1970)\citenamefont
  {Coester}, \citenamefont {Cohen}, \citenamefont {Day},\ and\ \citenamefont
  {Vincent}}]{coester1970prc}%
  \BibitemOpen
  \bibfield  {author} {\bibinfo {author} {\bibfnamefont {F.}~\bibnamefont
  {Coester}}, \bibinfo {author} {\bibfnamefont {S.}~\bibnamefont {Cohen}},
  \bibinfo {author} {\bibfnamefont {B.}~\bibnamefont {Day}},\ and\ \bibinfo
  {author} {\bibfnamefont {C.~M.}\ \bibnamefont {Vincent}},\ }\bibfield
  {title} {\bibinfo {title} {Variation in nuclear-matter binding energies with
  phase-shift-equivalent two-body potentials},\ }\href
  {https://doi.org/10.1103/PhysRevC.1.769} {\bibfield  {journal} {\bibinfo
  {journal} {Phys. Rev. C}\ }\textbf {\bibinfo {volume} {1}},\ \bibinfo {pages}
  {769} (\bibinfo {year} {1970})}\BibitemShut {NoStop}%
\bibitem [{\citenamefont {Day}\ and\ \citenamefont
  {Coester}(1976)}]{day1976prc}%
  \BibitemOpen
  \bibfield  {author} {\bibinfo {author} {\bibfnamefont {B.~D.}\ \bibnamefont
  {Day}}\ and\ \bibinfo {author} {\bibfnamefont {F.}~\bibnamefont {Coester}},\
  }\bibfield  {title} {\bibinfo {title} {Influence of virtual
  $\ensuremath{\Delta}$ states on the saturation properties of nuclear
  matter},\ }\href {https://doi.org/10.1103/PhysRevC.13.1720} {\bibfield
  {journal} {\bibinfo  {journal} {Phys. Rev. C}\ }\textbf {\bibinfo {volume}
  {13}},\ \bibinfo {pages} {1720} (\bibinfo {year} {1976})}\BibitemShut
  {NoStop}%
\bibitem [{\citenamefont {Day}(1981)}]{day1981prl}%
  \BibitemOpen
  \bibfield  {author} {\bibinfo {author} {\bibfnamefont {B.~D.}\ \bibnamefont
  {Day}},\ }\bibfield  {title} {\bibinfo {title} {Nuclear saturation from
  two-nucleon potentials},\ }\href {https://doi.org/10.1103/PhysRevLett.47.226}
  {\bibfield  {journal} {\bibinfo  {journal} {Phys. Rev. Lett.}\ }\textbf
  {\bibinfo {volume} {47}},\ \bibinfo {pages} {226} (\bibinfo {year}
  {1981})}\BibitemShut {NoStop}%
\bibitem [{\citenamefont {Li}\ \emph {et~al.}(2006)\citenamefont {Li},
  \citenamefont {Lombardo}, \citenamefont {Schulze}, \citenamefont {Zuo},
  \citenamefont {Chen},\ and\ \citenamefont {Ma}}]{li2006prc}%
  \BibitemOpen
  \bibfield  {author} {\bibinfo {author} {\bibfnamefont {Z.~H.}\ \bibnamefont
  {Li}}, \bibinfo {author} {\bibfnamefont {U.}~\bibnamefont {Lombardo}},
  \bibinfo {author} {\bibfnamefont {H.-J.}\ \bibnamefont {Schulze}}, \bibinfo
  {author} {\bibfnamefont {W.}~\bibnamefont {Zuo}}, \bibinfo {author}
  {\bibfnamefont {L.~W.}\ \bibnamefont {Chen}},\ and\ \bibinfo {author}
  {\bibfnamefont {H.~R.}\ \bibnamefont {Ma}},\ }\bibfield  {title} {\bibinfo
  {title} {Nuclear matter saturation point and symmetry energy with modern
  nucleon-nucleon potentials},\ }\href
  {https://doi.org/10.1103/PhysRevC.74.047304} {\bibfield  {journal} {\bibinfo
  {journal} {Phys. Rev. C}\ }\textbf {\bibinfo {volume} {74}},\ \bibinfo
  {pages} {047304} (\bibinfo {year} {2006})}\BibitemShut {NoStop}%
\bibitem [{\citenamefont {Bogner}\ \emph {et~al.}(2005)\citenamefont {Bogner},
  \citenamefont {Schwenk}, \citenamefont {Furnstahl},\ and\ \citenamefont
  {Nogga}}]{bogner2005npa}%
  \BibitemOpen
  \bibfield  {author} {\bibinfo {author} {\bibfnamefont {S.~K.}\ \bibnamefont
  {Bogner}}, \bibinfo {author} {\bibfnamefont {A.}~\bibnamefont {Schwenk}},
  \bibinfo {author} {\bibfnamefont {R.~J.}\ \bibnamefont {Furnstahl}},\ and\
  \bibinfo {author} {\bibfnamefont {A.}~\bibnamefont {Nogga}},\ }\bibfield
  {title} {\bibinfo {title} {Is nuclear matter perturbative with low-momentum
  interactions?},\ }\href {https://doi.org/10.1016/j.nuclphysa.2005.08.024}
  {\bibfield  {journal} {\bibinfo  {journal} {Nucl. Phys. A}\ }\textbf
  {\bibinfo {volume} {763}},\ \bibinfo {pages} {59} (\bibinfo {year}
  {2005})}\BibitemShut {NoStop}%
\bibitem [{\citenamefont {Horowitz}\ and\ \citenamefont
  {Serot}(1987)}]{horowitz1987npa}%
  \BibitemOpen
  \bibfield  {author} {\bibinfo {author} {\bibfnamefont {C.~J.}\ \bibnamefont
  {Horowitz}}\ and\ \bibinfo {author} {\bibfnamefont {B.~D.}\ \bibnamefont
  {Serot}},\ }\bibfield  {title} {\bibinfo {title} {The relativistic
  two-nucleon problem in nuclear matter},\ }\href
  {https://doi.org/10.106/0375-9474(87)90370-8} {\bibfield  {journal} {\bibinfo
   {journal} {Nuclear Physics A}\ }\textbf {\bibinfo {volume} {464}},\ \bibinfo
  {pages} {613} (\bibinfo {year} {1987})}\BibitemShut {NoStop}%
\bibitem [{\citenamefont {Sehn}\ \emph {et~al.}(1997)\citenamefont {Sehn},
  \citenamefont {Fuchs},\ and\ \citenamefont {Faessler}}]{sehn1997prc}%
  \BibitemOpen
  \bibfield  {author} {\bibinfo {author} {\bibfnamefont {L.}~\bibnamefont
  {Sehn}}, \bibinfo {author} {\bibfnamefont {C.}~\bibnamefont {Fuchs}},\ and\
  \bibinfo {author} {\bibfnamefont {A.}~\bibnamefont {Faessler}},\ }\bibfield
  {title} {\bibinfo {title} {Nucleon self-energy in the relativistic brueckner
  approach},\ }\href {https://doi.org/10.1103/PhysRevC.56.216} {\bibfield
  {journal} {\bibinfo  {journal} {Phys. Rev. C}\ }\textbf {\bibinfo {volume}
  {56}},\ \bibinfo {pages} {216} (\bibinfo {year} {1997})}\BibitemShut
  {NoStop}%
\bibitem [{\citenamefont {Fuchs}\ \emph {et~al.}(1998)\citenamefont {Fuchs},
  \citenamefont {Waindzoch}, \citenamefont {Faessler},\ and\ \citenamefont
  {Kosov}}]{fuchs1998prc}%
  \BibitemOpen
  \bibfield  {author} {\bibinfo {author} {\bibfnamefont {C.}~\bibnamefont
  {Fuchs}}, \bibinfo {author} {\bibfnamefont {T.}~\bibnamefont {Waindzoch}},
  \bibinfo {author} {\bibfnamefont {A.}~\bibnamefont {Faessler}},\ and\
  \bibinfo {author} {\bibfnamefont {D.~S.}\ \bibnamefont {Kosov}},\ }\bibfield
  {title} {\bibinfo {title} {Scalar and vector decomposition of the nucleon
  self-energy in the relativistic brueckner approach},\ }\href
  {https://doi.org/10.1103/PhysRevC.58.2022} {\bibfield  {journal} {\bibinfo
  {journal} {Phys. Rev. C}\ }\textbf {\bibinfo {volume} {58}},\ \bibinfo
  {pages} {2022} (\bibinfo {year} {1998})}\BibitemShut {NoStop}%
\bibitem [{\citenamefont {Ring}(1996)}]{ring1996pr}%
  \BibitemOpen
  \bibfield  {author} {\bibinfo {author} {\bibfnamefont {P.}~\bibnamefont
  {Ring}},\ }\bibfield  {title} {\bibinfo {title} {Relativistic mean field
  theory in finite nuclei},\ }\bibfield  {journal} {\bibinfo  {journal}
  {Progress in Particle and Nuclear Physics}\ }\textbf {\bibinfo {volume}
  {37}},\ \href {https://doi.org/10.1016/0146-6410(96)00054-3}
  {10.1016/0146-6410(96)00054-3} (\bibinfo {year} {1996})\BibitemShut {NoStop}%
\bibitem [{\citenamefont {Shen}\ \emph {et~al.}(2016)\citenamefont {Shen},
  \citenamefont {Hu}, \citenamefont {Liang}, \citenamefont {Meng},
  \citenamefont {Ring},\ and\ \citenamefont {Zhang}}]{shen2016cpl}%
  \BibitemOpen
  \bibfield  {author} {\bibinfo {author} {\bibfnamefont {S.~H.}\ \bibnamefont
  {Shen}}, \bibinfo {author} {\bibfnamefont {J.~N.}\ \bibnamefont {Hu}},
  \bibinfo {author} {\bibfnamefont {H.~Z.}\ \bibnamefont {Liang}}, \bibinfo
  {author} {\bibfnamefont {J.}~\bibnamefont {Meng}}, \bibinfo {author}
  {\bibfnamefont {P.}~\bibnamefont {Ring}},\ and\ \bibinfo {author}
  {\bibfnamefont {S.~Q.}\ \bibnamefont {Zhang}},\ }\bibfield  {title} {\bibinfo
  {title} {Relativistic brueckner—hartree—fock theory for finite nuclei},\
  }\href {https://doi.org/10.1088/0256-307X/33/10/102103} {\bibfield  {journal}
  {\bibinfo  {journal} {Chin. Phys. Lett.}\ }\textbf {\bibinfo {volume} {33}},\
  \bibinfo {pages} {102103} (\bibinfo {year} {2016})}\BibitemShut {NoStop}%
\bibitem [{\citenamefont {Shen}\ \emph {et~al.}(2017)\citenamefont {Shen},
  \citenamefont {Liang}, \citenamefont {Meng}, \citenamefont {Ring},\ and\
  \citenamefont {Zhang}}]{shen2017prc}%
  \BibitemOpen
  \bibfield  {author} {\bibinfo {author} {\bibfnamefont {S.~H.}\ \bibnamefont
  {Shen}}, \bibinfo {author} {\bibfnamefont {H.~Z.}\ \bibnamefont {Liang}},
  \bibinfo {author} {\bibfnamefont {J.}~\bibnamefont {Meng}}, \bibinfo {author}
  {\bibfnamefont {P.}~\bibnamefont {Ring}},\ and\ \bibinfo {author}
  {\bibfnamefont {S.~Q.}\ \bibnamefont {Zhang}},\ }\bibfield  {title} {\bibinfo
  {title} {Fully self-consistent relativistic brueckner-hartree-fock theory for
  finite nuclei},\ }\href {https://doi.org/10.1103/PhysRevC.96.014316}
  {\bibfield  {journal} {\bibinfo  {journal} {Phys. Rev. C}\ }\textbf {\bibinfo
  {volume} {96}},\ \bibinfo {pages} {014316} (\bibinfo {year}
  {2017})}\BibitemShut {NoStop}%
\bibitem [{\citenamefont {Wang}\ \emph {et~al.}(2021)\citenamefont {Wang},
  \citenamefont {Zhao}, \citenamefont {Ring},\ and\ \citenamefont
  {Meng}}]{wang2021prc}%
  \BibitemOpen
  \bibfield  {author} {\bibinfo {author} {\bibfnamefont {S.~B.}\ \bibnamefont
  {Wang}}, \bibinfo {author} {\bibfnamefont {Q.}~\bibnamefont {Zhao}}, \bibinfo
  {author} {\bibfnamefont {P.}~\bibnamefont {Ring}},\ and\ \bibinfo {author}
  {\bibfnamefont {J.}~\bibnamefont {Meng}},\ }\bibfield  {title} {\bibinfo
  {title} {Nuclear matter in relativistic brueckner-hartree-fock theory with
  bonn potential in the full dirac space},\ }\href
  {https://doi.org/10.1103/PhysRevC.103.054319} {\bibfield  {journal} {\bibinfo
   {journal} {Phys. Rev. C}\ }\textbf {\bibinfo {volume} {103}},\ \bibinfo
  {pages} {054319} (\bibinfo {year} {2021})}\BibitemShut {NoStop}%
\bibitem [{\citenamefont {Wang}\ \emph {et~al.}(2022)\citenamefont {Wang},
  \citenamefont {Tong},\ and\ \citenamefont {Wang}}]{wang2022prc}%
  \BibitemOpen
  \bibfield  {author} {\bibinfo {author} {\bibfnamefont {S.~B.}\ \bibnamefont
  {Wang}}, \bibinfo {author} {\bibfnamefont {H.}~\bibnamefont {Tong}},\ and\
  \bibinfo {author} {\bibfnamefont {C.~C.}\ \bibnamefont {Wang}},\ }\bibfield
  {title} {\bibinfo {title} {Nuclear matter within the continuous choice in the
  full dirac space},\ }\href {https://doi.org/10.1103/PhysRevC.105.054309}
  {\bibfield  {journal} {\bibinfo  {journal} {Phys. Rev. C}\ }\textbf {\bibinfo
  {volume} {105}},\ \bibinfo {pages} {054309} (\bibinfo {year}
  {2022})}\BibitemShut {NoStop}%
\bibitem [{\citenamefont {Stoks}\ \emph {et~al.}(1994)\citenamefont {Stoks},
  \citenamefont {Klomp}, \citenamefont {Terheggen},\ and\ \citenamefont
  {de~Swart}}]{stoks1994prc}%
  \BibitemOpen
  \bibfield  {author} {\bibinfo {author} {\bibfnamefont {V.~G.~J.}\
  \bibnamefont {Stoks}}, \bibinfo {author} {\bibfnamefont {R.~A.~M.}\
  \bibnamefont {Klomp}}, \bibinfo {author} {\bibfnamefont {C.~P.~F.}\
  \bibnamefont {Terheggen}},\ and\ \bibinfo {author} {\bibfnamefont {J.~J.}\
  \bibnamefont {de~Swart}},\ }\bibfield  {title} {\bibinfo {title}
  {Construction of high-quality nn potential models},\ }\href
  {https://doi.org/10.1103/PhysRevC.49.2950} {\bibfield  {journal} {\bibinfo
  {journal} {Phys. Rev. C}\ }\textbf {\bibinfo {volume} {49}},\ \bibinfo
  {pages} {2950} (\bibinfo {year} {1994})}\BibitemShut {NoStop}%
\bibitem [{\citenamefont {Dumbrajs}\ \emph {et~al.}(1983)\citenamefont
  {Dumbrajs}, \citenamefont {Koch}, \citenamefont {Pilkuhn}, \citenamefont
  {Oades}, \citenamefont {Behrens}, \citenamefont {de~Swart},\ and\
  \citenamefont {Kroll}}]{dumbrajs1983npb}%
  \BibitemOpen
  \bibfield  {author} {\bibinfo {author} {\bibfnamefont {O.}~\bibnamefont
  {Dumbrajs}}, \bibinfo {author} {\bibfnamefont {R.}~\bibnamefont {Koch}},
  \bibinfo {author} {\bibfnamefont {H.}~\bibnamefont {Pilkuhn}}, \bibinfo
  {author} {\bibfnamefont {G.~C.}\ \bibnamefont {Oades}}, \bibinfo {author}
  {\bibfnamefont {H.}~\bibnamefont {Behrens}}, \bibinfo {author} {\bibfnamefont
  {J.~J.}\ \bibnamefont {de~Swart}},\ and\ \bibinfo {author} {\bibfnamefont
  {P.}~\bibnamefont {Kroll}},\ }\bibfield  {title} {\bibinfo {title}
  {Compilation of coupling constants and low-energy parameters},\ }\href
  {https://doi.org/10.1016/0550-3213(83)90288-2} {\bibfield  {journal}
  {\bibinfo  {journal} {Nucl. Phys. B}\ }\textbf {\bibinfo {volume} {216}},\
  \bibinfo {pages} {277} (\bibinfo {year} {1983})}\BibitemShut {NoStop}%
\bibitem [{\citenamefont {Long}\ \emph {et~al.}(2006)\citenamefont {Long},
  \citenamefont {Giai},\ and\ \citenamefont {Meng}}]{long2006plb}%
  \BibitemOpen
  \bibfield  {author} {\bibinfo {author} {\bibfnamefont {W.~H.}\ \bibnamefont
  {Long}}, \bibinfo {author} {\bibfnamefont {N.~V.}\ \bibnamefont {Giai}},\
  and\ \bibinfo {author} {\bibfnamefont {J.}~\bibnamefont {Meng}},\ }\bibfield
  {title} {\bibinfo {title} {Density-dependent relativistic hartree–fock
  approach},\ }\href
  {https://doi.org/https://doi.org/10.1016/j.physletb.2006.07.064} {\bibfield
  {journal} {\bibinfo  {journal} {Phys. Lett. B}\ }\textbf {\bibinfo {volume}
  {640}},\ \bibinfo {pages} {150} (\bibinfo {year} {2006})}\BibitemShut
  {NoStop}%
\bibitem [{\citenamefont {Long}\ \emph {et~al.}(2007)\citenamefont {Long},
  \citenamefont {Sagawa}, \citenamefont {Giai},\ and\ \citenamefont
  {Meng}}]{long2007prc}%
  \BibitemOpen
  \bibfield  {author} {\bibinfo {author} {\bibfnamefont {W.~H.}\ \bibnamefont
  {Long}}, \bibinfo {author} {\bibfnamefont {H.}~\bibnamefont {Sagawa}},
  \bibinfo {author} {\bibfnamefont {N.~V.}\ \bibnamefont {Giai}},\ and\
  \bibinfo {author} {\bibfnamefont {J.}~\bibnamefont {Meng}},\ }\bibfield
  {title} {\bibinfo {title} {Shell structure and \ensuremath{\rho}-tensor
  correlations in density dependent relativistic hartree-fock theory},\ }\href
  {https://doi.org/10.1103/PhysRevC.76.034314} {\bibfield  {journal} {\bibinfo
  {journal} {Phys. Rev. C}\ }\textbf {\bibinfo {volume} {76}},\ \bibinfo
  {pages} {034314} (\bibinfo {year} {2007})}\BibitemShut {NoStop}%
\bibitem [{\citenamefont {Bogner}\ \emph {et~al.}(2007)\citenamefont {Bogner},
  \citenamefont {Furnstahl}, \citenamefont {Ramanan},\ and\ \citenamefont
  {Schwenk}}]{bogner2007npa}%
  \BibitemOpen
  \bibfield  {author} {\bibinfo {author} {\bibfnamefont {S.~K.}\ \bibnamefont
  {Bogner}}, \bibinfo {author} {\bibfnamefont {R.~J.}\ \bibnamefont
  {Furnstahl}}, \bibinfo {author} {\bibfnamefont {S.}~\bibnamefont {Ramanan}},\
  and\ \bibinfo {author} {\bibfnamefont {A.}~\bibnamefont {Schwenk}},\
  }\bibfield  {title} {\bibinfo {title} {Low-momentum interactions with smooth
  cutoffs},\ }\href {https://doi.org/10.1016/j.nuclphysa.2006.11.123}
  {\bibfield  {journal} {\bibinfo  {journal} {Nucl. Phys. A}\ }\textbf
  {\bibinfo {volume} {784}},\ \bibinfo {pages} {79} (\bibinfo {year}
  {2007})}\BibitemShut {NoStop}%
\bibitem [{\citenamefont {Wiringa}\ \emph {et~al.}(1995)\citenamefont
  {Wiringa}, \citenamefont {Stoks},\ and\ \citenamefont
  {Schiavilla}}]{wiringa1995prc}%
  \BibitemOpen
  \bibfield  {author} {\bibinfo {author} {\bibfnamefont {R.~B.}\ \bibnamefont
  {Wiringa}}, \bibinfo {author} {\bibfnamefont {V.~G.~J.}\ \bibnamefont
  {Stoks}},\ and\ \bibinfo {author} {\bibfnamefont {R.}~\bibnamefont
  {Schiavilla}},\ }\bibfield  {title} {\bibinfo {title} {Accurate
  nucleon-nucleon potential with charge-independence breaking},\ }\href
  {https://doi.org/10.1103/PhysRevC.51.38} {\bibfield  {journal} {\bibinfo
  {journal} {Phys. Rev. C}\ }\textbf {\bibinfo {volume} {51}},\ \bibinfo
  {pages} {38} (\bibinfo {year} {1995})}\BibitemShut {NoStop}%
\bibitem [{\citenamefont {Entem}\ \emph {et~al.}(2017)\citenamefont {Entem},
  \citenamefont {Machleidt},\ and\ \citenamefont {Nosyk}}]{entem2017prc}%
  \BibitemOpen
  \bibfield  {author} {\bibinfo {author} {\bibfnamefont {D.~R.}\ \bibnamefont
  {Entem}}, \bibinfo {author} {\bibfnamefont {R.}~\bibnamefont {Machleidt}},\
  and\ \bibinfo {author} {\bibfnamefont {Y.}~\bibnamefont {Nosyk}},\ }\bibfield
   {title} {\bibinfo {title} {High-quality two-nucleon potentials up to fifth
  order of the chiral expansion},\ }\href
  {https://doi.org/10.1103/PhysRevC.96.024004} {\bibfield  {journal} {\bibinfo
  {journal} {Phys. Rev. C}\ }\textbf {\bibinfo {volume} {96}},\ \bibinfo
  {pages} {024004} (\bibinfo {year} {2017})}\BibitemShut {NoStop}%
\bibitem [{\citenamefont {Brown}\ \emph {et~al.}(1969)\citenamefont {Brown},
  \citenamefont {Jackson},\ and\ \citenamefont {Kuo}}]{brown1969npa}%
  \BibitemOpen
  \bibfield  {author} {\bibinfo {author} {\bibfnamefont {G.~E.}\ \bibnamefont
  {Brown}}, \bibinfo {author} {\bibfnamefont {A.~D.}\ \bibnamefont {Jackson}},\
  and\ \bibinfo {author} {\bibfnamefont {T.~T.~S.}\ \bibnamefont {Kuo}},\
  }\bibfield  {title} {\bibinfo {title} {Nucleon-nucleon potential and minimal
  relativity},\ }\href {https://doi.org/10.1016/0375-9474(69)90549-1}
  {\bibfield  {journal} {\bibinfo  {journal} {Nucl. Phys. A}\ }\textbf
  {\bibinfo {volume} {133}},\ \bibinfo {pages} {481} (\bibinfo {year}
  {1969})}\BibitemShut {NoStop}%
\bibitem [{\citenamefont {Baldo}\ \emph {et~al.}(1991)\citenamefont {Baldo},
  \citenamefont {Bombaci}, \citenamefont {Ferreira}, \citenamefont
  {Giansiracusa},\ and\ \citenamefont {Lombardo}}]{baldo1991prc}%
  \BibitemOpen
  \bibfield  {author} {\bibinfo {author} {\bibfnamefont {M.}~\bibnamefont
  {Baldo}}, \bibinfo {author} {\bibfnamefont {I.}~\bibnamefont {Bombaci}},
  \bibinfo {author} {\bibfnamefont {L.~S.}\ \bibnamefont {Ferreira}}, \bibinfo
  {author} {\bibfnamefont {G.}~\bibnamefont {Giansiracusa}},\ and\ \bibinfo
  {author} {\bibfnamefont {U.}~\bibnamefont {Lombardo}},\ }\bibfield  {title}
  {\bibinfo {title} {Nuclear matter within the continuous choice},\ }\href
  {https://doi.org/10.1103/PhysRevC.43.2605} {\bibfield  {journal} {\bibinfo
  {journal} {Phys. Rev. C}\ }\textbf {\bibinfo {volume} {43}},\ \bibinfo
  {pages} {2605} (\bibinfo {year} {1991})}\BibitemShut {NoStop}%
\end{thebibliography}%

\end{document}